\documentclass[a4paper,fleqn]{cas-dc}

\usepackage[nameinlink,capitalize]{cleveref}
\usepackage[numbers,sort&compress]{natbib}
\usepackage{subcaption}
\usepackage{graphicx}
\usepackage{booktabs}
\usepackage{caption}
\usepackage{amsmath}
\usepackage{empheq}
\hypersetup{
    colorlinks=true,
    linkcolor=cyan,
    citecolor=cyan,
    urlcolor=cyan,
}

\def\tsc#1{\csdef{#1}{\textsc{\lowercase{#1}}\xspace}}
\tsc{WGM}
\tsc{QE}
\tsc{EP}
\tsc{PMS}
\tsc{BEC}
\tsc{DE}

\begin{document}
\let\WriteBookmarks\relax
\def\floatpagepagefraction{1}
\def\textpagefraction{.001}
\let\printorcid\relax
\shorttitle{}
\shortauthors{Wang et~al.}

\title [mode = title]{UF-AMA: A unified framework for cross-domain emotion recognition via adaptive multimodal alignment}

\author[1,3]{Zheng Wang}
\ead{eduwangz03@mail.ustc.edu.cn}

\credit{Writing – original draft, Visualization, Validation, Software, Methodology, Formal analysis, Data curation, Conceptualization}

\author[2]{Shuo Wang}
\ead{shuowang.edu@gmail.com}

\credit{Writing – review and editing, Visualization, Validation, Supervision, Project administration, Formal analysis}

\author[1,3]{Junhong Wang}\corref{cor1}
\ead{wjh007@iai.ustc.edu.cn}

\credit{Writing – review and editing, Resources, Investigation, Funding acquisition}

\affiliation[1]{organization={Institute of Advanced Technology, University of Science and Technology of China},
                city={Hefei},
                postcode={230026}, 
                country={China}}

\affiliation[2]{organization={Department of Electronic Engineering and Information Science, University of Science and Technology of China},
                city={Hefei},
                postcode={230026},
                country={China}}

\affiliation[3]{organization={Institute of Artificial Intelligence, Hefei Comprehensive National Science Center},
                city={Hefei},
                postcode={230026},
                country={China}}

\cortext[cor1]{Corresponding author}

\begin{abstract}
In recent years, emotion recognition based on physiological signals such as electroencephalogram (EEG) has gained considerable attention, as internal physiological data offer greater objectivity and reliability compared to external behavioral data like facial expressions. However, due to distribution shifts caused by individual and contextual differences, along with variations in sample quality across modalities, constructing a cross-domain multimodal emotion recognition model with high generalization and robustness remains a key challenge. In this study, we propose a Unified Framework with Adaptive Multimodal Alignment (UF-AMA) to address cross-subject and cross-session emotion recognition using multimodal physiological signals. First, we construct a cross-modal feature fusion network comprising Transformer encoders and multi-head cross-attention modules, enabling the deep integration of EEG signals and eye-tracking data. Subsequently, we introduce a confidence-aware screening mechanism that dynamically assesses the predictive reliability of each modality branch on target domain samples, partitions samples into different quality subsets, and accordingly applies global consistency alignment and cross-modal distillation. Finally, we propose a multi-level domain adaptation framework that jointly optimizes the marginal and conditional distributions of both local modality-specific and global fusion features, thereby reducing cross-domain distribution shifts at multiple granularities. Extensive experiments on the SEED and SEED-IV datasets demonstrate that UF-AMA achieves state-of-the-art (SOTA) performance in both cross-subject and cross-session tasks. The source code is available at: \href{https://github.com/BetterCoderLab/UF-AMA}{https://github.com/BetterCoderLab/UF-AMA}.
\end{abstract}


\maketitle

\section{Introduction}

Emotion is central to interpersonal communication, directly shaping language comprehension and expression, and thereby profoundly influencing social and cognitive development. Its influence also extends to decision-making, mental and physical health, and daily behavior \cite{Tyng2017}. Specifically, emotional changes trigger notable external behavioral alterations, including facial expressions, changes in prosody, body postures, as well as crying or sighing \cite{Ekman1993, Rawal2022, Rajapakshe2022}; emotions can also induce internal physiological responses, such as changes in heart rate, blood pressure, respiration rate, and the activation of emotion-related brain regions.

In recent years, with the maturation of data acquisition technologies like microphones and cameras, external behavioral data can be easily collected, yielding large-scale, learnable samples. Consequently, emotion recognition methods based on external behavioral data have seen considerable research and application. However, external behavioral data are susceptible to individual subjective concealment and environmental interference, making them inherently deceptive and fundamentally limiting their reliability. In contrast, physiological signals represented by EEG and ECG originate directly from the human autonomic and central nervous systems. They possess strong biological relevance and a low capacity for social concealment. As a result, they reflect authentic emotional fluctuations more objectively and reliably, being largely free from conscious control. Thus, they provide a more reliable biosensing foundation for emotion recognition and have attracted widespread attention.

As a spontaneous and difficult-to-disguise neurophysiological signal, EEG can objectively and real-timely map human emotional states \cite{Yang2017}. Numerous EEG-based emotion recognition studies have emerged in affective computing, constructing various high-performance classification models. For instance, Yin et al. \cite{Yin2021} integrated graph convolutional neural networks (GCNN) and long short-term memory (LSTM) networks, where the GCNN captures topological dependencies among EEG channels and the LSTM extracts temporal information, thereby enhancing emotion recognition. However, this method partially overlooks spatio-temporal evolution patterns and suffers from parameter inefficiency due to model stacking. To address these issues, An et al. \cite{An2025a} combined ConvLSTM modules with residual connections \cite{Shi2015} and a self-attention mechanism. The ConvLSTM extracts temporal correlations, spatial structures, and spectral information from spatio-temporal and spatial-spectral matrices, while the self-attention integrates cross-modal dependencies to enhance complementary information learning. Nevertheless, ConvLSTM is primarily designed for processing time-series data to capture spatial evolution patterns, whereas different frequency bands in spatial-spectral matrices exhibit parallel or coupled relationships without temporal order. Therefore, using ConvLSTM on spatial-spectral matrices is suboptimal, reflecting the difficulty of existing encoders to effectively and reasonably accommodate heterogeneous physiological features within a unified framework. Recently, the Transformer architecture \cite{Vaswani2017}, with its superior global dependency modeling capability, has emerged as a more universal alternative for emotion recognition. For example, Wang et al. \cite{Wang2022} used Transformer encoders to hierarchically integrate intra- and inter-regional spatial dependencies among brain regions, automatically enhancing emotionally critical features. Peng et al. \cite{Peng2023} introduced temporal relative encoding and channel attention into Transformer to better capture continuous emotional evolution and evaluate channel contributions. Beyond model architecture, Transformer has also been combined with federated learning to address privacy concerns. Gahlan et al. \cite{Gahlan2024} proposed an attention-based federated learning framework integrating Transformer and artificial neural networks for multimodal feature fusion.

Although EEG-based emotion recognition has advanced considerably, single-modality approaches remain inherently limited. Emotional states are highly complex to generate, and physiological signals are susceptible to noise during acquisition, making it difficult for a single modality to comprehensively and accurately characterize emotions \cite{Ji2023}. In contrast, multimodal physiological signal-based emotion recognition offers significant advantages: it compensates for single-modality limitations by integrating multi-perspective information, leverages heterogeneous complementary information across physiological representations to improve robustness, and captures intrinsic inter-modal correlations to significantly enhance recognition accuracy.

In recent years, multimodal emotion recognition has primarily adopted feature-level fusion for cross-modal complementarity. Although decision-level fusion fully considers modality heterogeneity, it often overlooks inter-modality complementarity and is thus less common. EEG signals typically play a central and dominant role in emotion representation. Accordingly, Li et al. \cite{Li2025} proposed a deep EEG-first multimodal method (DEMA), which uses deep multi-view convolutional neural networks to extract comprehensive multi-domain EEG features within an EEG-centered fusion framework. By introducing an affective influence matrix, they quantified EEG's impact on other physiological signals, achieving unified modeling of heterogeneous physiological representations. To address inconsistent emotion expression and information redundancy in multimodal fusion, Li et al. \cite{Li2024a} proposed a Transformer-based method with low-rank fusion \cite{Liu2018}, treating EEG as the primary modality. Cross-modal Transformers optimize auxiliary modality features to eliminate inter-modal conflicts, and low-rank fusion reduces redundancy, thereby substantially enhancing primary modality features. Furthermore, Gong et al. \cite{Gong2024} proposed an attention fusion network jointly considering multimodal heterogeneity and correlation. Dual-stream extractors decouple EEG and peripheral physiological signal (PPS) features, while intra-modality encoding and inter-modality fusion modules capture cross-modal correlations while preserving modality-specific heterogeneity. Building on this, Zhu et al. \cite{Zhu2025} proposed MF-net, which retains the dual-stream architecture with multi-scale optimization to extract emotional information from temporal slices of EEG and peripheral signals. A cross-modal local-global feature fusion module and an adaptive collaboration module dynamically allocate modality weights to enhance critical information expression.

However, physiological signals exhibit significant individual variability across subjects due to factors such as age, gender, cultural background, and psychological state. Even within the same subject, physiological representations vary with psychological state fluctuations or environmental changes, displaying notable non-stationarity and cross-session distribution shifts. To address this, researchers have explored unsupervised domain adaptation (UDA), which transfers knowledge from a well-labeled source domain to an unlabeled or sparsely labeled target domain to enhance model generalization. Sartipi et al. \cite{Sartipi2023} combined a Transformer with adversarial discriminative domain adaptation to alleviate performance degradation caused by individual differences. Ai et al. \cite{Ai2025} proposed a multi-source domain independent matching method, treating each subject as an independent source domain and using multi-channel adversarial training for alignment. Building on their earlier framework, An et al. \cite{An2025a} introduced a local-global domain adversarial discriminator to synergistically reduce distribution bias across multi-scale feature spaces. Beyond adversarial learning, statistical metric alignment is also commonly adopted. Gretton et al. \cite{Gretton2012} used maximum mean discrepancy (MMD) \cite{Ouyang2021} to minimize the Hilbert space distance between source and target feature distributions. Miao et al. \cite{Miao2023} proposed MLDA, which uses MMD for marginal distribution alignment while introducing a correlation domain adaptation strategy to alleviate conditional distribution discrepancies. An et al. \cite{An2025b} proposed STCBI-Nets with a joint optimized adaptive domain alignment strategy (JOADAS), incorporating an adaptive class-center alignment loss based on global adversarial learning to enhance intra-class aggregation and cross-subject generalization.

Currently, existing multimodal methods primarily focus on mining cross-modal complementary information, while subject distribution alignment is predominantly performed in the global feature space of the fusion branch. For instance, Tang et al. \cite{Tang2024} proposed RHPR-Net, which introduces a random contrastive loss in the global fusion feature space to suppress individual differences. Xu et al. \cite{Xu2024} proposed LAFDA-Net, which integrates a domain adaptation branch into the fused global feature space, jointly using covariance distance and domain classifier consistency constraints to reduce cross-domain distribution shifts. However, since different modalities exhibit different drift characteristics during cross-domain transfer, single global alignment often masks modality-specific distribution biases. Therefore, collaborative alignment of multi-level structural representations is necessary. For example, Jiménez-Guarneros et al. \cite{Jimenez-Guarneros2024a} proposed CFDA-CSF, which performs fine-grained structural distribution alignment within each modality in addition to overall fused feature alignment. They further introduced a cross-modal alignment loss to exploit inter-modal correlations by constraining representation consistency across modalities in a common feature space.

\begin{figure*}
	\centering
	\includegraphics[width=\textwidth]{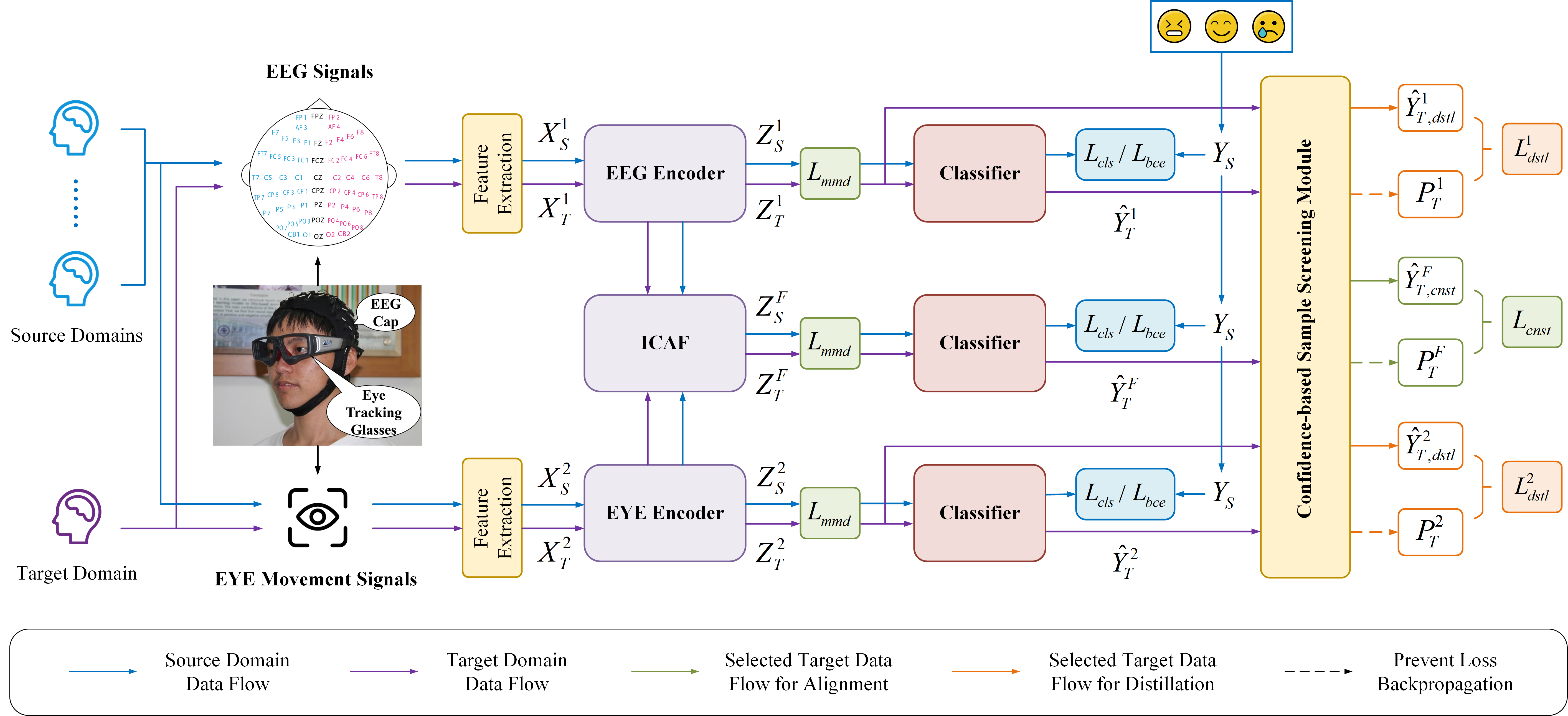}
	\caption{Overall flowchart of the proposed UF-AMA.}
	\label{FIG:1}
\end{figure*}

Although multi-level collaborative alignment improves generalization, existing methods remain largely limited to coarse-grained marginal distribution alignment, neglecting semantic-guided conditional distribution alignment and thus failing to capture fine-grained emotional category boundaries. Moreover, existing semantic guidance strategies often overlook modality-level quality differences in the target domain, where noise from low-quality modalities can lead to error accumulation \cite{Liu2025}. Therefore, constructing a framework that balances coarse-to-fine distribution alignment with modality quality awareness is crucial. 

To address these challenges, we propose a unified framework with adaptive multimodal alignment (UF-AMA). Our framework employs a cross-modal feature fusion network built upon unified Transformer encoding and multi-head cross-attention mechanisms, which jointly integrate coarse-grained marginal alignment with fine-grained semantic alignment. To mitigate error accumulation caused by modality quality discrepancies, we introduce a confidence-aware sample screening mechanism into the fine-grained semantic alignment process. Additionally, we assign different weights to this alignment across different training stages, ensuring that the network prioritizes learning effective and reliable representations from the source domain in the early stage, thereby preventing error accumulation due to unreliable sample screening in the initial phase. The main contributions of this study are as follows:

\begin{enumerate}[(1)]
    \item We propose an end-to-end training framework with a multi-level domain adaptation strategy, comprising a multi-task classification structure on the source domain and a local-global distribution alignment framework, thereby ensuring discriminative representations of both local modality-specific and global fusion features while effectively reducing cross-domain distribution shifts at multiple granularities.
    \item We propose a confidence-aware sample screening mechanism that dynamically evaluates the prediction confidence and consistency of target domain samples across modalities, selectively stratifying them into high-quality, conflicting, and partially low-quality subsets, which are then further processed accordingly to suppress pseudo-label noise accumulation.
    \item Extensive evaluations on the SEED and SEED-IV datasets under cross-subject and cross-session settings demonstrate that our method achieves state-of-the-art (SOTA) performance in recognition accuracy and generalization.
\end{enumerate}

The remainder of this paper is structured as follows: Section 2 elaborates on the model architecture and implementation details of UF-AMA; Section 3 presents experimental settings and result analysis on the SEED and SEED-IV datasets; Section 4 provides an in-depth discussion of the experimental results; Section 5 concludes the paper. Data availability is mentioned at the end.

\section{Methodology}

\cref{FIG:1} depicts the overall framework of UF-AMA, where we construct a multi-level deep learning pipeline comprising the following core components: a cross-modal feature fusion network, a coarse-grained distribution alignment module, and a confidence-based fine-grained semantic alignment module. The cross-modal feature fusion network consists of intra-modality Transformer encoders and an inter-modal cross-attention fusion (ICAF) module for feature extraction and deep semantic aggregation. Additionally, a local–global collaborative multi-task learning mechanism is incorporated to further refine the discriminative boundaries of the feature space through multi-granularity supervision. The coarse-grained distribution alignment module employs maximum mean discrepancy (MMD) to achieve inter-domain distribution alignment. The confidence-based fine-grained semantic alignment module integrates a sample selection mechanism, global consistency alignment, and local cross-modal distillation, enabling accurate calibration of target domain samples of varying quality through dynamic supervisory signals.

\subsection{Feature extraction}

To extract discriminative emotional representations from raw EEG and eye movement data, we construct a standardized preprocessing and feature extraction pipeline. The EEG preprocessing pipeline includes downsampling, bandpass filtering, sliding-window segmentation, and frequency-band feature extraction. Considering the computational burden caused by the high sampling rate, the EEG signals are downsampled from 1000 Hz to 200 Hz. A bandpass filter between 1 Hz and 75 Hz is applied to remove noise and artifacts. The signals are segmented into non-overlapping 4-second epochs and decomposed into five standard frequency bands: $\delta$ (1-4 Hz), $\theta$ (4-8 Hz), $\alpha$ (8-14 Hz), $\beta$ (14-31 Hz), and $\gamma$ (31-50 Hz), each corresponding to distinct physiological activities of brain regions. For each band, we extract differential entropy (DE) as the core feature, as it demonstrates strong robustness against noise in non-stationary signals such as EEG \cite{Duan2013}. As a continuous analog of discrete Shannon entropy, its calculation is defined as follows:
\begin{equation}
    H(\mathbf{x}) = -\int p(\mathbf{x}) \log p(\mathbf{x}) \, d\mathbf{x}
    \label{EQ:DE}
\end{equation}
where $p(\mathbf{x})$ denotes the probability density function of the amplitude distribution within a given frequency band, and $H(\mathbf{x})$ accordingly represents the differential entropy of that band. For eye movement data, we further compute statistical features of pupil diameter, fixations, saccades, and blinks to compensate for the limitations of EEG in capturing arousal and other dimensions.

The DE features of all channels across all frequency bands for one EEG segment are organized into a structured feature matrix, whose dimensions correspond to the number of channels and frequency bands. These segment-level matrices are aggregated across all trials for a single subject to form the preprocessed EEG feature data. Combined with the statistical features of the eye movement modality, this preprocessing pipeline provides a robust data foundation that enables precise localization of neural patterns associated with specific emotional states and delivers discriminative semantic input for emotion classification.

\subsection{Network architecture}

The cross-modal feature fusion network consists of three core components: intra-modality encoders, an inter-modal cross-attention fusion (ICAF) module, and multi-task classification heads. Each intra-modality encoder is built by stacking multiple Transformer encoder blocks, which integrate feed-forward networks, multi-head self-attention, and residual connections to map preprocessed features into a high-dimensional latent space. The ICAF module combines linear projections, multi-head cross-attention, and residual connections to capture semantic dependencies across modalities by computing their dynamic correlations. The multi-task classification heads consist of fully connected layers that produce predictions from the fused features and each modality's features, and jointly optimize cross-entropy loss and binary cross-entropy loss to ensure strong inter-class discriminability and intra-class semantic consistency.

\subsubsection{Intra-modality encoder}

The Transformer architecture \cite{Vaswani2017}, with its self-attention mechanism, flexibly accommodates heterogeneous feature representations across modalities while simultaneously offering superior global modeling capacity. Therefore, we adopt it as the feature extraction backbone for both EEG and eye movement data. Specifically, we construct two independent encoder branches for processing EEG and eye movement data, respectively. During training, each encoder receives input exclusively from its corresponding modality, and the branches remain mutually independent, accepting no external data flow from the other modality. This preserves intra-modality information purity and avoids prematurely introducing cross-modal interference.

Each modality encoder is composed of several stacked Transformer encoder blocks, each containing a multi-head self-attention (MHSA) computation unit and a feed-forward network. The multi-head self-attention unit comprises several attention heads. The attention calculation formula can be expressed as follows:
\begin{equation}
    \operatorname{Attention}(\mathbf{Q}, \mathbf{K}, \mathbf{V}) = \operatorname{softmax}\left(\frac{\mathbf{Q}\mathbf{K}^T}{\sqrt{d_k}}\right)\mathbf{V}
    \label{EQ:attention}
\end{equation}
where $\mathbf{Q}$, $\mathbf{K}$, and $\mathbf{V}$ denote the query, key, and value matrices, respectively, and $d_k$ is the dimensionality of the key vectors. Subsequently, the outputs of all attention heads are concatenated and linearly projected, followed by a residual connection. The feed-forward network (FFN) consists of two fully connected layers with a ReLU activation in between, also followed by a residual connection. For the input $\mathbf{X}^m$ of modality $m$, the forward propagation through the encoder block is formulated as follows:
\begin{equation}
    \left\{
    \begin{aligned}
        & \bar{\mathbf{X}}_m^{(t)} = \operatorname{LN}(\mathbf{Z}_m^{(t-1)}), \quad \mathbf{Z}_m^{(0)} = \mathbf{X}^m \\
        & \text{SA}_{m}^{h(t)} = \operatorname{Attention}(\bar{\mathbf{X}}_m^{(t)} \mathbf{W}_h^Q, \bar{\mathbf{X}}_m^{(t)} \mathbf{W}_h^K, \bar{\mathbf{X}}_m^{(t)} \mathbf{W}_h^V) \\
        & \text{MHSA}_m^{(t)} = \operatorname{Concat}(\text{SA}_{m}^{1(t)}, \dots, \text{SA}_{m}^{H(t)}) \mathbf{W}^O \\
        & \bar{\mathbf{Z}}_m^{(t)} = \mathbf{Z}_m^{(t-1)} + \text{MHSA}_m^{(t)} \\
        & \mathbf{Z}_m^{(t)} = \bar{\mathbf{Z}}_m^{(t)} + \operatorname{FFN}(\operatorname{LN}(\bar{\mathbf{Z}}_m^{(t)}))
    \end{aligned}
    \right.
\end{equation}
where $m \in \{1, 2\}$ denotes the modality index (EEG and eye movement, respectively), $t = 1, \ldots, T$ is the encoder block index, and $\operatorname{LN}$ denotes layer normalization.

Through the aforementioned deep encoding process, each modality encoder can extract highly semantically discriminative features, thereby providing a high-quality semantic representation foundation for the subsequent cross-modal fusion process.

\subsubsection{Inter-modality cross-attention fusion}

\begin{figure}
	\centering
	\includegraphics[width=\columnwidth]{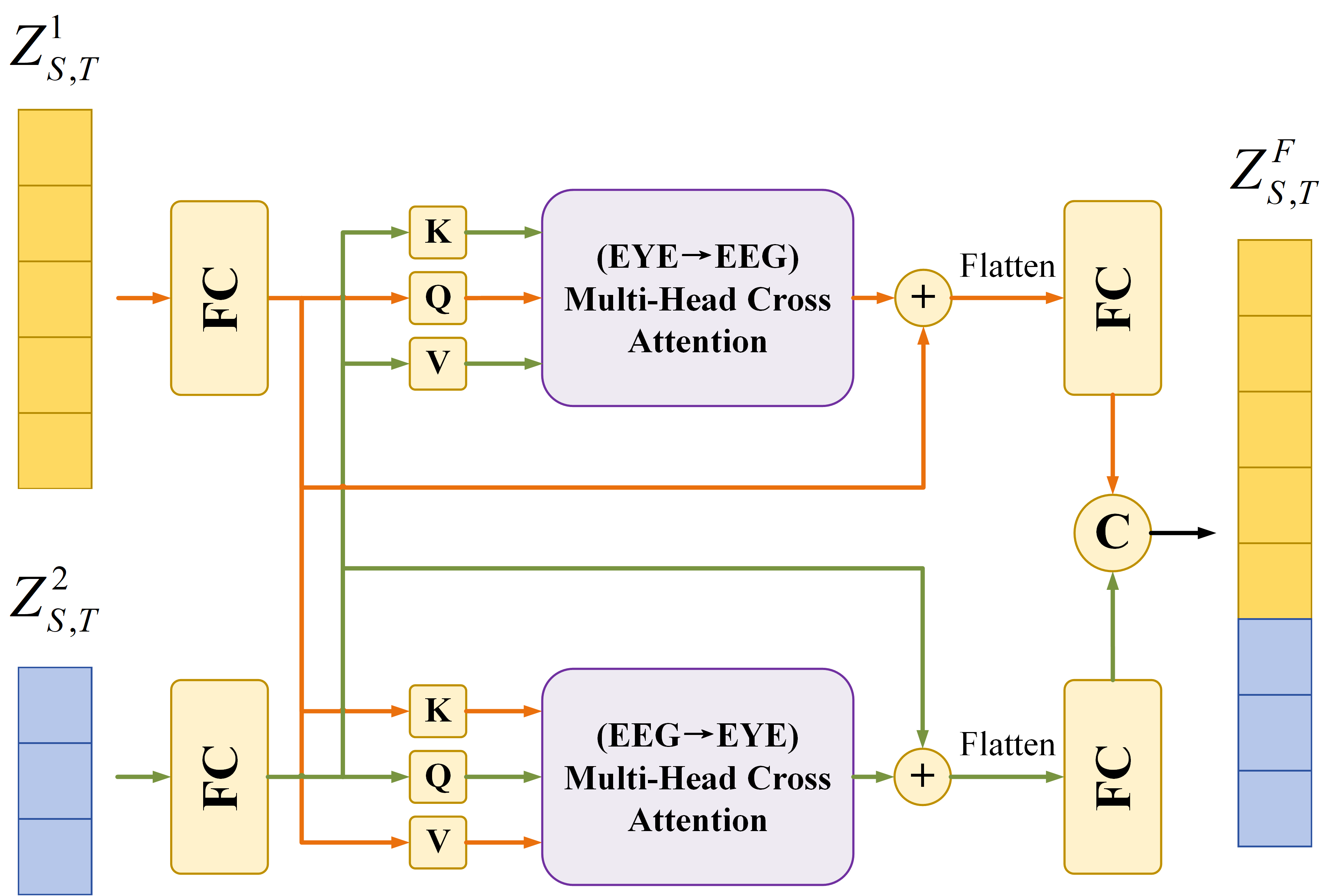}
	\caption{Architecture of the inter-modality cross-attention fusion (ICAF) module.}
	\label{FIG:2}
\end{figure}

The EEG modality captures emotional arousal through frequency-band responses, while the eye movement modality captures behavioral reactions elicited by visual stimuli, endowing the two modalities with natural complementarity. To effectively exploit this synergy, we design a fusion module centered on multi-head cross-attention (MHCA) \cite{Lin2022}, as illustrated in \cref{FIG:2}. 

Specifically, the encoded features from each modality are first projected into a shared high-dimensional space via linear layers. A bidirectional cross-attention mechanism is then established, where each modality symmetrically serves as the query to attend to the other and residual connections are employed throughout to preserve original information. This mechanism adaptively enhances shared emotional patterns while suppressing modality-specific noise. Finally, the outputs from both branches are concatenated to form the fused feature representation.

The multi-head cross-attention mechanism in the fusion pipeline is implemented by two multi-head cross-attention computation units. Each unit contains several attention heads, and the outputs of all attention heads are concatenated and integrated through a linear projection, and the computation is finalized with a residual connection. Given the encoded features $\mathbf{Z}_m$ and $\mathbf{Z}_n$ from two modalities, the integration process of the attention heads in the cross-attention mechanism can be expressed as follows:
\begin{equation}
    \left\{
    \begin{aligned}
        & \bar{\mathbf{Z}}_m = \operatorname{FC}(\operatorname{LN}(\mathbf{Z}_m)), \quad \bar{\mathbf{Z}}_n = \operatorname{FC}(\operatorname{LN}(\mathbf{Z}_n)) \\
        & \text{CA}_{n \to m}^{h} = \operatorname{Attention}(\bar{\mathbf{Z}}_m \mathbf{W}_h^Q, \bar{\mathbf{Z}}_n \mathbf{W}_h^K, \bar{\mathbf{Z}}_n \mathbf{W}_h^V) \\
        & \text{MHCA}_{n \to m} = \operatorname{Concat}(\text{CA}_{n \to m}^{1}, \dots, \text{CA}_{n \to m}^{H}) \mathbf{W}^O \\
        & \bar{\mathbf{Z}}_{n \to m} = \bar{\mathbf{Z}}_m + \text{MHCA}_{n \to m}
    \end{aligned}
    \right.
\end{equation}
where the subscripts $m$ and $n$ index the two modalities, $\operatorname{FC}$ denotes a fully connected layer.

Through the above process, the EEG and eye movement modalities each obtain their corresponding modal fusion patterns $\bar{\mathbf{Z}}_{P \to E}$ and $\bar{\mathbf{Z}}_{E \to P}$, where $E$ and $P$ denote EEG and eye movement. Finally, the pattern mapping and concatenation process is formulated as follows:
\begin{empheq}{align}
    & {\mathbf{Z}}_{fusion} = \operatorname{Concat}(\operatorname{FC}(\bar{\mathbf{Z}}_{P \to E}), \operatorname{FC}(\bar{\mathbf{Z}}_{E \to P}))
\end{empheq}

Based on the fusion process described above, we can effectively integrate emotional cues from EEG and eye movement data, compensating for the limitations of single modalities in emotional expression, and ultimately obtain a fused feature with strong complementarity, providing a high-quality input foundation for the classifiers.

\subsubsection{Classification}

Having obtained the fused features along with the encoded features from each individual modality, we employ a local–global collaborative training strategy to enhance the discriminative power of the learned representations. By fully mining multi-granularity semantic information in the source domain, this strategy establishes a strong prior for the subsequent fine-grained semantic alignment in the target domain, thereby substantially improving the robustness and accuracy of semantic anchoring during cross-domain adaptation. 

Accordingly, we jointly train the model by applying a cross-entropy loss to each modality-specific branch as well as the fusion branch. Furthermore, we introduce a binary cross-entropy (BCE) loss to enhance intra-class compactness, enabling the model to cope with intra-class feature distribution variations and thus improving the clustering quality of the encoded features from each modality encoder and the fused features in the feature space. For two source domain samples with feature representations $\mathbf{x}_i$ and $\mathbf{x}_j$ and their corresponding labels $\mathbf{y}_i$ and $\mathbf{y}_j$ within the same modality, we first define an indicator function that identifies whether the two samples belong to the same class:
\begin{equation}
    \mathbf{Y}_{ij} = 
    \begin{cases}
        1, & \mathbf{y}_i = \mathbf{y}_j \\
        0, & \mathbf{y}_i \neq \mathbf{y}_j
    \end{cases}
    \label{EQ:indicator}
\end{equation}

Next, we employ cosine similarity to measure the semantic affinity between $\mathbf{x}_i$ and $\mathbf{x}_j$, and normalize the resulting similarity score to the $(0, 1)$ interval:
\begin{equation}
    \mathbf{S}_{ij} = \frac{1}{2}\cos(\mathbf{x}_i, \mathbf{x}_j) + \frac{1}{2}
    \label{EQ:similarity}
\end{equation}

Finally, by minimizing the binary cross-entropy between the predicted similarity distribution $\mathbf{S}_{ij}$ and the ground-truth indicator matrix $\mathbf{Y}_{ij}$, the model is enforced to pull feature representations of same-class samples closer together while pushing those of different classes apart. The overall BCE loss is formulated as follows:
\begin{multline}
    \mathcal{L}_{\text{bce}} = \frac{1}{N(N-1)} \sum_{i \neq j} 
    \bigl[ \mathbf{Y}_{ij} \log \mathbf{S}_{ij} \\
    + (1 - \mathbf{Y}_{ij}) \log (1 - \mathbf{S}_{ij}) \bigr]
    \label{EQ:bce_loss}
\end{multline}

\subsection{Coarse-grained distribution alignment}

To reduce the marginal distribution discrepancy between the source and target domains, we introduce the Maximum Mean Discrepancy (MMD) loss for coarse-grained distribution alignment. The core idea of MMD is to map the features of different domains into a reproducing kernel Hilbert space (RKHS) and measure the distribution discrepancy by computing the distance between the mean embeddings of the two domains. During training, we minimize this distance in the kernel space, thereby forcing the model to extract domain-invariant feature representations and achieving a basic local--global alignment of marginal distributions. Specifically, the coarse-grained distribution alignment loss based on MMD is formulated as follows:
\begin{equation}
    \mathcal{L}_{\text{mmd}} = \left\Vert \frac{1}{N_S} \sum_{i=1}^{N_S} \phi(\mathbf{x}_i^S) - \frac{1}{N_T} \sum_{j=1}^{N_T} \phi(\mathbf{x}_j^T) \right\Vert_{\mathcal{H}}
    \label{EQ:mmd}
\end{equation}
where $\mathbf{x}_i^S$ and $\mathbf{x}_j^T$ denote the feature representations of the source domain and target domain samples, $N_S$ and $N_T$ are the corresponding batch sizes, $\phi(\cdot)$ is the implicit feature mapping induced by a kernel function, and $\Vert \cdot \Vert_{\mathcal{H}}$ denotes the norm in the RKHS $\mathcal{H}$.

In practice, we adopt a multi-kernel MMD (MK-MMD) strategy to enhance the representation capacity of the kernel space. Specifically, \cref{EQ:mmd} is expanded via a convex combination of multiple Gaussian kernels, ultimately yielding the following computable form:
\begin{equation}
    \begin{aligned}
        \mathcal{L}_{\text{mmd}} = & \frac{1}{N_S^2} \sum_{i=1}^{N_S} \sum_{j=1}^{N_S} \kappa(\mathbf{x}_i^S, \mathbf{x}_j^S) \\
        & + \frac{1}{N_T^2} \sum_{i=1}^{N_T} \sum_{j=1}^{N_T} \kappa(\mathbf{x}_i^T, \mathbf{x}_j^T) \\
        & - \frac{2}{N_S N_T} \sum_{i=1}^{N_S} \sum_{j=1}^{N_T} \kappa(\mathbf{x}_i^S, \mathbf{x}_j^T)
    \end{aligned}
    \label{EQ:mkmmd}
\end{equation}
where $\kappa(\cdot, \cdot) = \sum_{u=0}^{K-1} k_u(\cdot, \cdot)$ denotes the multi-kernel function with $K$ Gaussian kernels, and $k_u$ is the $u$-th Gaussian radial basis function (RBF) kernel.

\subsection{Fine-grained semantic alignment}

\begin{figure*}
	\centering
	\includegraphics[width=\textwidth]{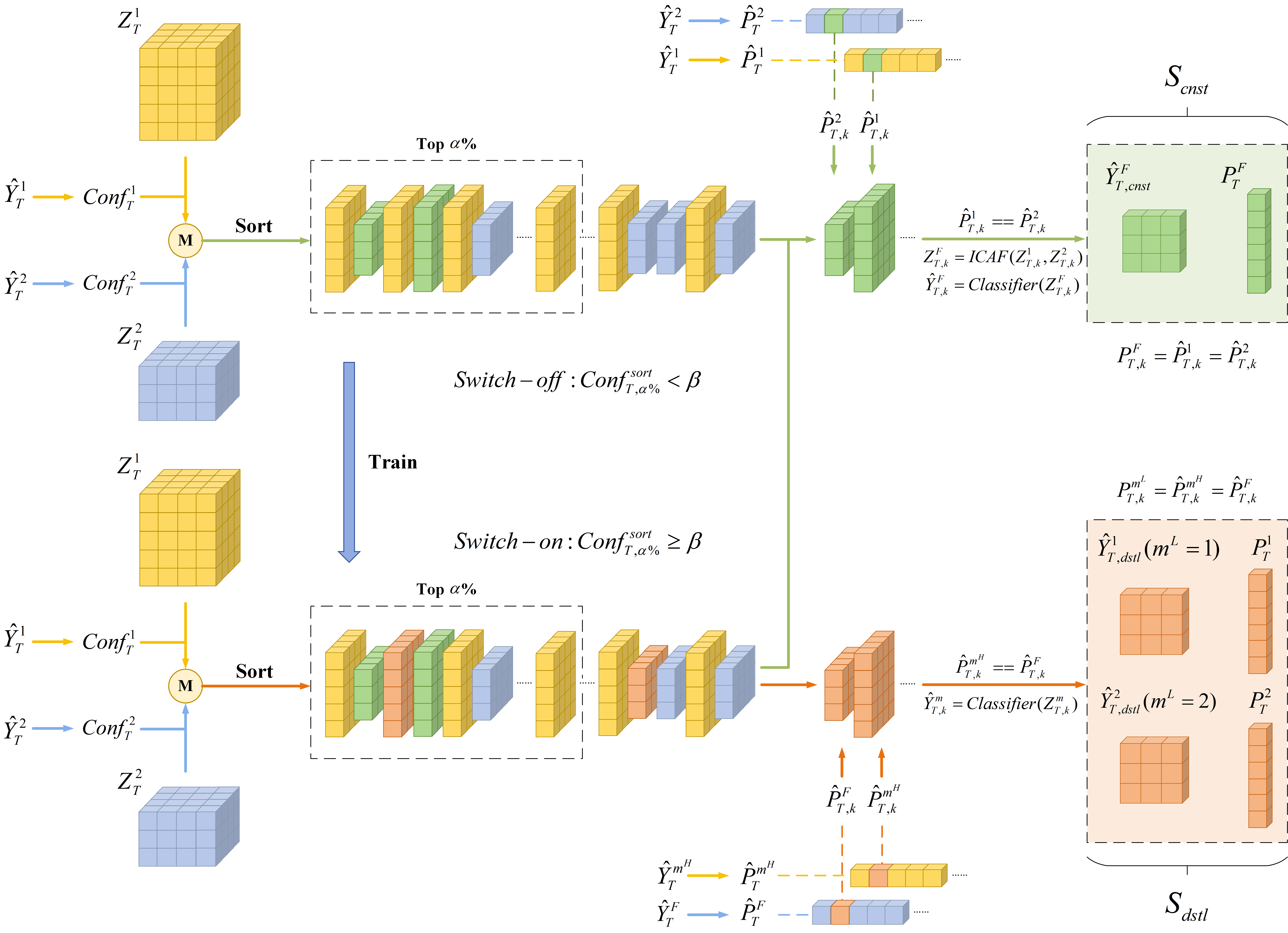}
	\caption{Workflow of the confidence-based sample screening mechanism. Following threshold-based screening, the retained samples are categorized into two main types: high-confidence consistent samples (green dashed boxes), which are sent to the global consistency alignment module; and partially low-quality samples (orange dashed boxes), which are routed to the local cross-modal distillation module.}
	\label{FIG:3}
\end{figure*}

Coarse-grained marginal distribution alignment reduces domain discrepancy but ignores class-conditional ones, risking semantic misalignment in the target domain. Our fine-grained semantic alignment thus introduces sample-level and class-level constraints to calibrate target domain feature boundaries, compensating for the inadequate inter-class discrimination caused by coarse-grained alignment. To effectively integrate this fine-grained module without incurring early-stage noise, we design a staged training procedure. In the early stage, with unstable target pseudo-labels, the model relies on source supervision and coarse-grained alignment to learn domain invariant features. As semantic learning deepens, we introduce the fine-grained semantic alignment module. At this stage, reliable pseudo-labels from confidence screening enable us to introduce global consistency alignment and local cross-modal distillation, which avoids error propagation while significantly improving generalization and robustness by precisely fine-tuning hard samples.

\subsubsection{Confidence-based sample screening}

In multi-modal emotion recognition tasks, data quality often varies significantly across modalities. In the source domain, modality specific noise can be implicitly reweighted through supervised gradients. However, the target domain, which lacks annotations, is far more sensitive to low quality modalities. Direct multi-modal fusion in this setting can cause noise to diffuse through the feature space, leading to erroneous pseudo labels. These errors then propagate inaccurate information and accumulate over time, ultimately limiting the model's overall recognition performance. To enable more refined feature processing, we design a confidence-based sample screening mechanism.

Specifically, as the target domain samples pass through the cross-modal feature fusion network, the fused features $\mathbf{Z}_T^F$ and the encoded features $\mathbf{Z}_T^m$ from each modality branch are output separately. The latter are subsequently fed into their respective emotion classifiers to produce predicted probability distributions $\hat{\mathbf{Y}}_T^m$, from which we obtain the pseudo-labels $\hat{\mathbf{P}}_T^m$ and the confidence scores $\mathbf{Conf}_T^m$. For the $k$-th sample in modality $m$, its pseudo-label $\hat{\mathbf{P}}_{k}^m$ is taken as the class with the highest predicted probability, and its confidence score $\mathbf{Conf}_k^m$ is defined as that maximum probability.

We gather modality-specific encoded features, confidence scores, and pseudo-labels for all target domain samples across all modalities to form a candidate modality sample set $\mathcal{X}$. To prioritize the selection of reliable and high-quality samples, we sort the elements of $\mathcal{X}$ in descending order of confidence score, producing a sorted modality sample sequence $\mathcal{X}_{\text{sort}}$. We then introduce a confidence ranking threshold $\alpha$ and retain the modality-level samples whose confidence ranks fall within the top $\alpha\%$ of the sorted sequence, forming the high-quality modality sample set $\mathcal{X}_{\text{high}}$. The screening process is expressed as follows:
\begin{equation}
    \mathcal{X}_{\text{high}} = \left\{ (\mathbf{Z}_{i, k}^{m}, \hat{\mathbf{P}}_{i, k}^{m}) \;\middle|\; i \leq \lceil \alpha\% \cdot |\mathcal{X}_{\text{sort}}| \rceil \right\}
    \label{EQ:screening}
\end{equation}
where $i$ denotes the $i$-th element in $\mathcal{X}_{\text{sort}}$, $k$ denotes the corresponding sample index, $m$ denotes the corresponding modality, and $\alpha$ is the confidence ranking threshold.

\cref{FIG:3} illustrates the workflow of the sample selection mechanism. After obtaining the high-quality modality sample set $\mathcal{X}_{\text{high}}$, the available target domain samples can be categorized into three distinct types:

\begin{enumerate}[(1)]
    \item High-confidence consistent samples: the encoded features of all modalities pass the screening and are present in $\mathcal{X}_{\text{high}}$, and the pseudo-labels across modalities remain strictly consistent. These samples serve as reliable semantic anchors and can directly participate in global consistency alignment.
    
    \item Semantically conflicting samples: the encoded features of all modalities  pass the screening and appear in $\mathcal{X}_{\text{high}}$, but there exists a fundamental discrepancy among their modality pseudo-labels. Thus, they are discarded to avoid the negative impact of conflicting information.
    
    \item Partially low-quality samples: only the encoded features of one modality pass the screening and are present in $\mathcal{X}_{\text{high}}$, while the other modality does not. These samples are primarily used for local cross-modal distillation, with the aim of leveraging the predictive information from the high-quality modality to perform feature calibration on the low-quality one.
\end{enumerate}

\subsubsection{Global consistency alignment}

The primary objective of applying global consistency alignment to high-confidence consistent samples is to correct the semantic deviation of their cross-modal fused features from the source domain priors. Specifically, the cross-domain shifts in modality-specific conditional distributions induce a deviation of the fused feature space from the semantic distribution of the corresponding source domain space. This deviation causes the prediction results of certain fused features to become inconsistent with the source domain priors, thereby giving rise to semantic discrimination bias. By imposing global consistency constraints, we semantically align the fused feature space with that of the source domain. This ensures that high-confidence consistent samples maintain consistency with the source domain not only at the modality-level semantic distribution but also in the post-fusion feature space, thereby achieving intrinsic cross-domain feature alignment.

We employ the cross-entropy loss to implement global consistency alignment. The entire global consistency alignment process can be expressed as follows:
\begin{equation}
    \mathcal{L}_{\text{cnst}} = -\frac{1}{|\mathcal{S}_{\text{cnst}}|} \sum_{k \in \mathcal{S}_{\text{cnst}}} \mathbf{P}_{k}^{F} \log \hat{\mathbf{Y}}_{cnst,k}^F
    \label{EQ:global_consistency}
\end{equation}
where $\mathcal{S}_{\text{cnst}}$ denotes the set of high-confidence consistent samples identified by the screening mechanism, $\mathbf{P}_{k}^{F}$ denotes the agreed-upon label across modalities for the $k$-th high-confidence consistent sample, $\mathbf{P}_{k}^{F} = \hat{\mathbf{P}}_{k}^{1} = \hat{\mathbf{P}}_{k}^{2}$, and $\hat{\mathbf{Y}}_{cnst,k}^F$ denotes the predicted probability distribution of the final fused feature from $\mathcal{S}_{\text{cnst}}$.

\subsubsection{Local cross-modal distillation}

For partially low-quality samples that do not satisfy the high-confidence consistency condition, we further introduce a local cross-modal distillation mechanism to fully exploit the semantic information masked by noise and strengthen the discriminative capacity of each single-modality branch. In this module, the modality with high confidence is assigned the role of teacher, while the modality with low confidence caused by missing information or noise interference is treated as the student. Furthermore, to ensure the reliability of the distillation process, we introduce a confidence gating threshold \(\beta\). As shown in \cref{FIG:3}, the local cross-modal distillation module is only activated when the lowest confidence score among the selected samples exceeds this threshold. Formally, let \(Conf_{\alpha\%}^{\text{sort}}\) denote the confidence score of the sample at the \(\alpha\%\) position in \(\mathcal{X}_{\text{sort}}\). The distillation process is activated only if \(Conf_{\alpha\%}^{\text{sort}} \geq \beta\).

To ensure the accuracy of knowledge transfer, we only retain samples where the fusion network's prediction result is consistent with the pseudo-label of the teacher modality. Subsequently, we calculate the distillation loss between the prediction vector of the student modality branch and the pseudo-label of the fusion branch for these retained samples. The cross-modal distillation process for modality $m^L$ can be expressed as follows:
\begin{equation}
    \mathcal{L}_{\text{dstl}}^{m^L} = -\frac{1}{|\mathcal{S}_{\text{dstl}}|} \sum_{k \in \mathcal{S}_{\text{dstl}}} \mathbf{P}_{k}^{m^L} \log \hat{\mathbf{Y}}_{dstl,k}^{m^L}
    \label{EQ:distillation}
\end{equation}
where $\mathcal{S}_{\text{dstl}}$ is the set of retained samples for which the fusion prediction is consistent with the teacher modality pseudo-label, $m^L$ denotes the student modality, $\mathbf{P}_{k}^{m^L}$ denotes the pseudo-label of the teacher modality $m^H$ and the fusion branch, $\mathbf{P}_{k}^{m^L} = \hat{\mathbf{P}}_{k}^{m^H} = \hat{\mathbf{P}}_{k}^{F}$, and $\hat{\mathbf{Y}}_{dstl,k}^{m^L}$ is the predicted probability distribution of the student modality branch.

\subsection{Overall learning objectives}

For the entire framework, we define the overall loss function as the weighted sum of the classification loss ($\mathcal{L}_{\text{cls}}$), intra-class compactness loss ($\mathcal{L}_{\text{bce}}$), marginal alignment loss ($\mathcal{L}_{\text{mmd}}$), global consistency loss ($\mathcal{L}_{\text{cnst}}$), and cross-modal distillation loss ($\mathcal{L}_{\text{dstl}}$). The expression is as follows:
\begin{equation}
    \mathcal{L}_{\text{total}} = \mathcal{L}_{\text{cls}} + \mathcal{L}_{\text{bce}} + \delta \mathcal{L}_{\text{mmd}} + \theta \mathcal{L}_{\text{cnst}} + \gamma \mathcal{L}_{\text{dstl}}
    \label{EQ:total_loss}
\end{equation}
where except for $\mathcal{L}_{\text{cnst}}$ and $\mathcal{L}_{\text{dstl}}$, each loss component is composed of a weighted combination of the EEG branch, eye movement branch, and fusion branch, following a distribution ratio where the EEG and eye movement branches each account for $1/4$ of the weight, and the fusion branch accounts for $1/2$. For $\mathcal{L}_{\text{dstl}}$, the two modality branches share equal weights. In the early stage of training, when the reliability of the target domain pseudo-labels is still low, we dynamically adjust the weighting coefficient $\theta$ of the global consistency loss $\mathcal{L}_{\text{cnst}}$ based on the convergence status of the classification loss. This dynamic weighting mechanism is formulated as follows:
\begin{equation}
    \theta = 
    \begin{cases}
    0.5, & \text{if } \mathcal{L}_{\text{cls}} \leq 0.1 \\
    0.2, & \text{if } 0.1 < \mathcal{L}_{\text{cls}} < 0.2 \\
    0.1, & \text{otherwise}
    \end{cases}
    \label{EQ:cnst_weight}
\end{equation}

Furthermore, for the cross-modal distillation loss, our motivation is to fine-tune the original discriminative boundaries through auxiliary supervisory signals, thereby enhancing the model's generalization ability. Accordingly, we assign it a relatively small fixed weighting coefficient \(\gamma\), aiming to prevent possible deviation or structural damage to the discriminative boundaries, ensuring the model's stability during target domain adaptation.

\section{Experiments and results}

\subsection{Dataset description}

In this study, we adopted the SEED and SEED-IV datasets constructed by the BCMI Lab at Shanghai Jiao Tong University as the performance evaluation benchmarks.

The SEED dataset \cite{Zheng2015} encompasses EEG signals from 15 subjects and eye movement data from 12 of these subjects. Each subject participated in three experimental sessions, with each session containing 15 independent trials. Each trial consisted of a 5-second start prompt, approximately 4 minutes of film clip viewing, 45 seconds of self-assessment feedback, and 15 seconds of rest. The experiment induced positive, neutral, and negative emotional states by presenting 15 carefully selected film clips. No two film clips targeting the same emotion were shown consecutively within a single experimental session. During the experiment, EEG signals were recorded using a 62-channel electrode cap at a sampling rate of 1000 Hz, and the signals were subsequently downsampled to 200 Hz after preprocessing. The electrode layout strictly followed the international 10-20 system standard. Eye movement data were synchronously collected using SMI eye-tracking glasses.

The SEED-IV dataset \cite{Zheng2018}, as an extension of the SEED dataset, encompasses EEG signals and eye movement data from 15 subjects. Each subject participated in three experimental sessions, with each session expanded to 24 independent trials. Each trial consisted of a 5-second start prompt, approximately 2 minutes of film clip viewing, and 45 seconds of self-assessment feedback. The SEED-IV dataset further refines the emotion categories to include four emotions: happy, sad, fearful, and neutral. The elicitation materials remained carefully selected film clips, but the number was expanded from the original 15 to 72 clips. During the experiment, EEG signals were similarly recorded using a 62-channel electrode cap at a sampling rate of 1000 Hz, and the signals were subsequently downsampled to 200 Hz after preprocessing. The electrode layout also strictly followed the international 10-20 system standard. Eye movement data were synchronously collected using SMI eye-tracking glasses. \cref{TAB:1} presents the details of the two datasets.

\begin{table}[htbp]
    \centering
    \captionsetup{singlelinecheck=off, font={bf, small}, labelsep=period}
    \caption{Comparison of SEED and SEED-IV datasets.}
    \label{TAB:1}
    \setlength{\tabcolsep}{8pt} 
    \renewcommand{\arraystretch}{1.3} 
    \footnotesize
    \begin{tabular*}{\columnwidth}{@{\hspace{8pt}}l @{\extracolsep{\fill}} cc @{\hspace{8pt}}}
        \toprule
        Details & SEED & SEED-IV \\
        \midrule
        Participants               & 12          & 15 \\
        Emotional stimuli          & Movie clips & Movie clips \\
        Sessions per participant   & 3           & 3 \\
        Trials per session         & 15          & 24 \\
        Physiological modalities   & 2           & 2 \\
        EEG channels               & 62          & 62 \\
        Sampling rate              & 200 Hz      & 200 Hz \\
        Target emotional classes   & Three       & Four \\
        \bottomrule
    \end{tabular*}
\end{table}

\subsection{Experimental settings}
\label{SUBSEC:3.2}

\begin{table}[htbp]
    \centering
    \captionsetup{singlelinecheck=off, font={bf, small}, labelsep=period}
    \caption{Hyper-parameters settings for SEED and SEED-IV datasets.}
    \label{TAB:2}
    \setlength{\tabcolsep}{8pt} 
    \renewcommand{\arraystretch}{1.3} 
    \footnotesize
    \begin{tabular*}{\columnwidth}{@{\hspace{8pt}} l @{\extracolsep{\fill}} cc @{\hspace{8pt}}}
        \toprule
        Details & SEED & SEED-IV \\
        \midrule
        Optimizer               & Adam          & Adam \\
        Batch size              & 32            & 32 \\
        Training epochs         & 400           & 400 \\
        Learning rate           & 5e-4          & 5e-4 \\
        Confidence ranking threshold: $\alpha$ & 0.5   & 0.5 \\
        Confidence gating threshold: $\beta$    & 0.9   & 0.9 \\
        Number of subjects (cross-subject)   & 12    & 15 \\
        Number of sessions (cross-session)   & 3     & 3 \\
        \bottomrule
    \end{tabular*}
\end{table}

All experiments were conducted on an NVIDIA RTX 4090 GPU with 24 GB of VRAM. The software environment consisted of Python 3.8, PyTorch 1.11.0 for model development and training, and CUDA 11.3 for GPU acceleration. During training, the model adopted the Adam optimizer \cite{Kingma2014} for parameter updates, with the learning rate set to 5e-4, the batch size uniformly set to 32, and the number of training epochs set to 400 to ensure sufficient convergence. For the fine-grained semantic alignment module within the proposed framework, the confidence ranking threshold $\alpha$ and the confidence gating threshold $\beta$ were set to 0.5 and 0.9, respectively. Furthermore, the dropout rate in the multi-task classification heads was set to 0.5 to mitigate potential overfitting. \cref{TAB:2} details the comprehensive hyper-parameters of the cross-modal feature fusion network. Within the cross-modal feature fusion network, the intra-modal encoders employ a stack of three Transformer encoder blocks to capture subtle emotional differences in a deep semantic mapping space, and the number of attention heads for both the multi-head self-attention and multi-head cross-attention mechanisms was set to 2, striking a balance between feature extraction capability and overfitting suppression.

The experimental evaluation primarily used accuracy as the core metric, with the standard deviation as a supplementary measure. To systematically and comprehensively assess the classification performance and generalization ability of the proposed UF-AMA, we designed two types of experimental protocols for evaluation:

\begin{enumerate}[(1)]
    \item Cross-subject experiment: In this type of experiment, we employed the leave-one-subject-out cross-validation method for evaluation. That is, the data from one subject were iteratively selected as the target domain dataset, while the data from the remaining subjects served as the source domain dataset. After each round of training, the target domain dataset was used as the test set to evaluate the performance of the current model, thereby examining the model's generalization ability to the physiological emotional patterns of unseen individuals. Finally, the average accuracy across all subjects was calculated to obtain the final experimental result. For the SEED dataset, the sample size for the target domain was set as 1 subject × 3 sessions × 3394 samples, and the sample size for the source domain was set as 11 subjects × 3 sessions × 3394 samples. For the SEED-IV dataset, the total film duration varied across the three sessions, leading to an inconsistent sample size per experimental session. Consequently, the sample size for the target domain was set as 1 subject × (3435 + 3364 + 3324) samples, and the sample size for the source domain was set as 14 subjects × (3435 + 3364 + 3324) samples.
    \item Cross-session experiment: In this type of experiment, we employed the leave-one-session-out cross-validation method for evaluation. That is, the data from all subjects in one experimental session were iteratively selected as the target domain dataset, while the data from all subjects in the remaining experimental sessions served as the source domain dataset. After each round of training, the target domain dataset was used as the test set to evaluate the model's performance, thereby examining the model's generalization ability across sessions. Finally, the average accuracy across all sessions was calculated to obtain the final experimental result. For the SEED dataset, the sample size for the target domain was set as 1 session × 12 subjects × 3394 samples, and the sample size for the source domain was set as 2 sessions × 12 subjects × 3394 samples. For the SEED-IV dataset, the sample sizes for the target and source domains were determined on a case-by-case basis.
\end{enumerate}

\subsection{Experiments on the SEED dataset}

We evaluated UF-AMA on the SEED dataset under both cross-subject and cross-session protocols. The proposed method was systematically compared against a series of state-of-the-art unimodal and multimodal baselines.

The unimodal emotion recognition baselines included: ACRNN \cite{Tao2020}, which employs channel attention combined with CNN and self-attention RNN to extract spatio-temporal discriminative features; DA-CapsNet \cite{Liu2024}, which guides capsules to focus on complementary local features through a diversity enhancement mechanism; PR-PL \cite{Zhou2023}, which encodes semantic structures using prototype representations and combines pairwise learning to mitigate label noise; TSSSA-Net \cite{Guo2025}, which introduces a synchronous attention encoder to balance the contributions of time, frequency, and spatial domain features; LGDAAN-Nets \cite{An2025a}, which combine ConvLSTMs to extract multi-domain features and utilize local-global discriminators to capture domain-invariant features; STCBI-Nets \cite{An2025b}, which model brain region interactions through multi-head cross-attention and introduce contrastive learning to enhance synergistic representations; FMLAN \cite{Yu2025}, which utilizes a multi-subnet mutual learning mechanism to achieve cross-domain information complementarity; SDC-Net \cite{Tang2025}, which achieves unsupervised domain adaptation through within-trial data augmentation and statistical alignment.

The multimodal emotion recognition baselines included: CoDF-Net \cite{Gong2024a}, which maximizes modality correlations and utilizes a broad learning system for decision fusion to enhance robustness; CFDA-CSF \cite{Jimenez-Guarneros2024a}, which performs coarse-grained and fine-grained distribution alignment globally and inter-modally to learn a separable feature space; MSBLS \cite{Gong2023}, which mines multimodal residual information using a stacked broad learning system; and LAFDA-Net \cite{Xu2024}, which extracts features using parallel branches and reduces computational complexity and overfitting risk through low-rank fusion branches.

\begin{table}[htbp]
    \centering
    \captionsetup{singlelinecheck=off, font={bf, small}, labelsep=period}
    \caption{Performance comparison of various recognition methods in terms of accuracy and standard deviation on the SEED dataset.}
    \label{TAB:3}
    \setlength{\tabcolsep}{6pt} 
    \footnotesize 
    \renewcommand{\arraystretch}{1.3} 
    \begin{tabular*}{\columnwidth}{@{\hspace{6pt}} l l c c @{\hspace{6pt}}}
        \toprule
        Years & Method & \multicolumn{2}{c}{Accuracy/Std (\%)} \\ 
        \cmidrule(lr){3-4} & & Cross-subject & Cross-session \\
        \midrule
        2023 & ACRNN \cite{Tao2020}       & 76.30 $\pm$ 08.10 & 78.49 $\pm$ 09.87 \\
        2024 & DA-CapsNet \cite{Liu2024}  & 84.63 $\pm$ 09.09 & 88.18 $\pm$ 06.21 \\
        2024 & PR-PL \cite{Zhou2023}      & 85.56 $\pm$ 04.78 & {\textbf{93.18 $\pm$ 06.55}} \\
        2025 & TSSSA-Net \cite{Guo2025}   & 85.11 $\pm$ 10.31 & -- \\
        2025 & LGDAAN-Nets \cite{An2025a} & 89.09 $\pm$ 07.76 & 91.25 $\pm$ 05.87 \\
        2025 & STCBI-Nets \cite{An2025b}  & 90.21 $\pm$ 06.32 & 91.87 $\pm$ 05.53 \\
        2025 & FMLAN \cite{Yu2025}        & 90.96 $\pm$ 07.05 & 91.94 $\pm$ 07.73 \\
        2025 & SDC-Net \cite{Tang2025}    & \underline{91.85 $\pm$ 05.98} & -- \\
        2023 & CoDF-Net \cite{Gong2024a}  & 87.04 $\pm$ 06.58 & -- \\
        2024 & CFDA-CSF \cite{Jimenez-Guarneros2024a}   & 90.04 $\pm$ 05.46 & -- \\
        2024 & MSBLS \cite{Gong2023}      & 90.16 $\pm$ 06.40 & -- \\
        2024 & LAFDA-Net \cite{Xu2024}    & 91.70 $\pm$ 03.04 & -- \\
        \midrule
        2026 & UF-AMA & {\textbf{94.53 $\pm$ 07.11}} & \underline{92.39 $\pm$ 02.00} \\
        \bottomrule
    \end{tabular*}
    \flushleft{\scriptsize The best experimental results are bolded, and the second-best experimental results are underlined.}
\end{table}

\cref{TAB:3} presents the performance comparison of different methods on the SEED dataset. The experimental results show that the proposed UF-AMA achieved a classification accuracy of 94.53\% in the cross-subject experiment, ranking first among all compared algorithms. Concurrently, it achieved an accuracy of 92.39\% in the cross-session experiment, ranking second among all algorithms, surpassed only by PR-PL (93.18\%). Compared with the baseline models, UF-AMA demonstrated clear performance advantages in both cross-subject and cross-session experiments. In the cross-subject experiment, its accuracy was 2.68\% higher than that of the state-of-the-art model SDC-Net (91.85\%) and 2.83\% higher than LAFDA-Net (91.70\%), the best-performing model in multimodal tasks. In the cross-session experiment, its accuracy was 1.14\% higher than that of the recent representative high-performance model LGDAAN-Nets (91.25\%). Although the average accuracy of UF-AMA was 0.79\% lower than that of PR-PL in the cross-session experiment, its standard deviation was merely 2.00, significantly lower than the latter's 6.55. This indicates that UF-AMA exhibits considerably stronger robustness when facing session-to-session fluctuations, thus yielding more stable and reliable model outputs.

\begin{table*}[htbp]
    \centering
    \captionsetup{singlelinecheck=off, font={bf, small}, labelsep=period}
    \caption{Performance comparison for each subject and session in terms of accuracy and standard deviation on the SEED dataset.}
    \label{TAB:4}
    \renewcommand{\arraystretch}{1.3} 
    \setlength{\tabcolsep}{3pt} 
    \scriptsize
    \begin{tabular*}{\textwidth}{@{\hspace{3pt}} l @{\extracolsep{\fill}} cccccccccccc | cc @{\hspace{3pt}}}
        \toprule
        \textbf{Method} & \textbf{sub1} & \textbf{sub2} & \textbf{sub3} & \textbf{sub4} & \textbf{sub5} & \textbf{sub6} & \textbf{sub7} & \textbf{sub8} & \textbf{sub9} & \textbf{sub10} & \textbf{sub11} & \textbf{sub12} & \textbf{Mean} & \textbf{Std} \\
        \midrule
        
        \multicolumn{12}{l}{\textbf{Session 1}} & \multicolumn{1}{c}{} & \\
        CFDA-CSF & 81.00 & 92.43 & 83.80 & 93.82 & 91.42 & 97.50 & 98.60 & 92.28 & 100.00 & 88.20 & 97.55 & 100.00 & 93.05 & 6.22 \\
        LAFDA-Net & 83.23 & \textbf{94.97} & 86.11 & \textbf{96.45} & 93.94 & 100.00 & 98.74 & 94.84 & 100.00 & \textbf{90.63} & \textbf{98.21} & 100.00 & \textbf{94.75} & \textbf{5.31} \\
        UF-AMA & \textbf{93.11} & 88.24 & \textbf{93.11} & 87.65 & \textbf{100.00} & \textbf{100.00} & \textbf{100.00} & \textbf{100.00} & \textbf{100.00} & 67.70 & 86.10 & \textbf{100.00} & 92.99 & 9.28 \\
        \midrule
        
        \multicolumn{12}{l}{\textbf{Session 2}} & \multicolumn{1}{c}{} & \\
        CFDA-CSF & 77.20 & 93.85 & 100.00 & 82.70 & 100.00 & 84.00 & 100.00 & 100.00 & 84.05 & 58.47 & 69.10 & 81.07 & 85.87 & 13.52 \\
        LAFDA-Net & 79.45 & 96.59 & \textbf{100.00} & 85.11 & 100.00 & \textbf{86.45} & 100.00 & 98.91 & 86.55 & 60.17 & 71.11 & 83.43 & 87.31 & 12.20 \\
        UF-AMA & \textbf{93.94} & \textbf{100.00} & 92.16 & \textbf{100.00} & \textbf{100.00} & 86.22 & \textbf{100.00} & \textbf{100.00} & \textbf{100.00} & \textbf{86.10} & \textbf{93.11} & \textbf{87.05} & \textbf{94.88} & \textbf{5.64} \\
        \midrule
        
        \multicolumn{12}{l}{\textbf{Session 3}} & \multicolumn{1}{c}{} & \\
        CFDA-CSF & 78.13 & 94.55 & 93.38 & \textbf{92.20} & 100.00 & 100.00 & 76.95 & 80.07 & 89.67 & 95.07 & \textbf{94.33} & 100.00 & 91.20 & 8.39 \\
        LAFDA-Net & 91.75 & \textbf{100.00} & \textbf{94.28} & 87.87 & 100.00 & 100.00 & 78.61 & 91.87 & 90.24 & 92.07 & 89.61 & 100.00 & 93.03 & 6.15 \\
        UF-AMA & \textbf{100.00} & 88.24 & 93.59 & 87.41 & \textbf{100.00} & \textbf{100.00} & \textbf{100.00} & \textbf{93.11} & \textbf{100.00} & \textbf{100.00} & 86.10 & \textbf{100.00} & \textbf{95.70} & \textbf{5.45} \\
        \bottomrule
    \end{tabular*}
    \flushleft{\scriptsize The best experimental results are bolded.}
\end{table*}

\begin{table*}[htbp]
    \centering
    \captionsetup{singlelinecheck=off, font={bf, small}, labelsep=period}
    \caption{Performance comparison for each subject and session in terms of accuracy and standard deviation on the SEED-IV dataset.}
    \label{TAB:5}
    \renewcommand{\arraystretch}{1.3} 
    \setlength{\tabcolsep}{3pt}
    \scriptsize
    \begin{tabular*}{\textwidth}{@{\hspace{3pt}} l @{\extracolsep{\fill}} ccccccccccccccc | cc @{\hspace{3pt}}}
        \toprule
        \textbf{Method} & \textbf{sub1} & \textbf{sub2} & \textbf{sub3} & \textbf{sub4} & \textbf{sub5} & \textbf{sub6} & \textbf{sub7} & \textbf{sub8} & \textbf{sub9} & \textbf{sub10} & \textbf{sub11} & \textbf{sub12} & \textbf{sub13} & \textbf{sub14} & \textbf{sub15} & \textbf{Mean} & \textbf{Std} \\
        \midrule
        
        \multicolumn{15}{l}{\textbf{Session 1}} & \multicolumn{1}{c}{} & \\
        CFDA-CSF & 79.21 & 100.00 & 84.69 & 84.13 & 91.20 & 57.60 & 93.70 & 89.02 & 84.95 & 89.21 & 94.64 & 78.84 & 100.00 & 71.23 & 87.36 & 85.72 & 11.02 \\
        LAFDA-Net & 82.78 & 88.51 & \textbf{100.00} & \textbf{87.92} & 95.31 & 60.25 & 97.92 & \textbf{93.03} & \textbf{88.78} & 93.23 & \textbf{98.91} & 82.39 & \textbf{100.00} & 74.44 & \textbf{91.35} & 88.99 & 10.40 \\
        UF-AMA & \textbf{87.43} & \textbf{100.00} & 91.07 & 78.50 & \textbf{96.94} & \textbf{76.73} & \textbf{100.00} & 92.71 & 87.90 & \textbf{94.24} & 97.30 & \textbf{92.60} & 95.06 & \textbf{90.83} & 78.61 & \textbf{90.66} & \textbf{7.32} \\
        \midrule
        
        \multicolumn{15}{l}{\textbf{Session 2}} & \multicolumn{1}{c}{} & \\
        CFDA-CSF & 88.63 & 96.25 & 92.79 & 94.18 & 89.35 & 83.99 & 90.63 & 100.00 & 78.02 & 89.30 & 83.56 & 80.02 & 94.86 & 83.92 & 98.56 & 89.60 & 6.65 \\
        LAFDA-Net & 90.01 & 97.75 & 94.23 & \textbf{95.65} & 90.74 & 85.35 & 92.04 & 100.00 & 79.23 & 90.69 & 84.86 & 81.27 & 96.34 & 85.23 & \textbf{100.00} & 90.89 & 6.36 \\
        UF-AMA & \textbf{92.91} & \textbf{98.56} & \textbf{98.80} & 94.59 & \textbf{98.20} & \textbf{93.39} & \textbf{95.19} & \textbf{100.00} & \textbf{86.66} & \textbf{92.91} & \textbf{97.12} & \textbf{90.14} & \textbf{100.00} & \textbf{87.26} & 94.23 & \textbf{94.66} & \textbf{4.13} \\
        \midrule
        
        \multicolumn{15}{l}{\textbf{Session 3}} & \multicolumn{1}{c}{} & \\
        CFDA-CSF & 88.02 & 92.48 & 91.69 & 93.00 & 100.00 & 83.59 & 96.15 & 89.63 & 62.99 & 96.14 & 63.25 & 86.07 & 89.42 & 89.46 & 81.30 & 86.88 & 10.79 \\
        LAFDA-Net & 89.75 & 94.37 & 93.49 & 94.83 & 100.00 & 85.23 & \textbf{98.04} & 91.39 & 64.23 & 98.03 & 64.49 & 87.76 & 91.18 & 91.22 & \textbf{82.99} & 88.47 & 9.47 \\
        UF-AMA & \textbf{97.69} & \textbf{99.88} & \textbf{98.30} & \textbf{97.69} & \textbf{100.00} & \textbf{93.31} & 97.20 & \textbf{97.20} & \textbf{78.95} & \textbf{100.00} & \textbf{76.64} & \textbf{97.69} & \textbf{93.31} & \textbf{93.43} & 71.61 & \textbf{92.86} & \textbf{8.93} \\
        \bottomrule
    \end{tabular*}
    \flushleft{\scriptsize The best experimental results are bolded.}
\end{table*}

As shown in \cref{TAB:4}, the proposed UF-AMA demonstrated strong performance across multiple sessions of the SEED dataset. UF-AMA achieved the highest accuracy on Subject 1, Subject 3, Subjects 5-9, and Subject 12 in Session 1, attaining a 100.00\% recognition rate for 6 of these subjects. In Session 2 and Session 3, the leading advantage of the proposed method was further highlighted. From a global perspective, UF-AMA achieved an average accuracy of 94.88\% in Session 2, representing improvements of 9.01\% and 7.57\% over CFDA-CSF and LAFDA-Net, respectively. In Session 3, UF-AMA attained the highest average accuracy in the entire table at 95.70\%, representing improvements of 4.50\% and 2.67\% over CFDA-CSF and LAFDA-Net, respectively. Furthermore, the standard deviations of UF-AMA in Session 2 and Session 3 were only 5.64 and 5.45, respectively, which were significantly lower than the 13.52 and 8.39 of CFDA-CSF and the 12.20 and 6.15 of LAFDA-Net, once again demonstrating its superior stability when processing EEG signals and eye movement data.

\begin{figure*}[tbp]
    \centering
    \begin{subfigure}[b]{0.6\textwidth}
        \raggedright
        \includegraphics[height=6cm]{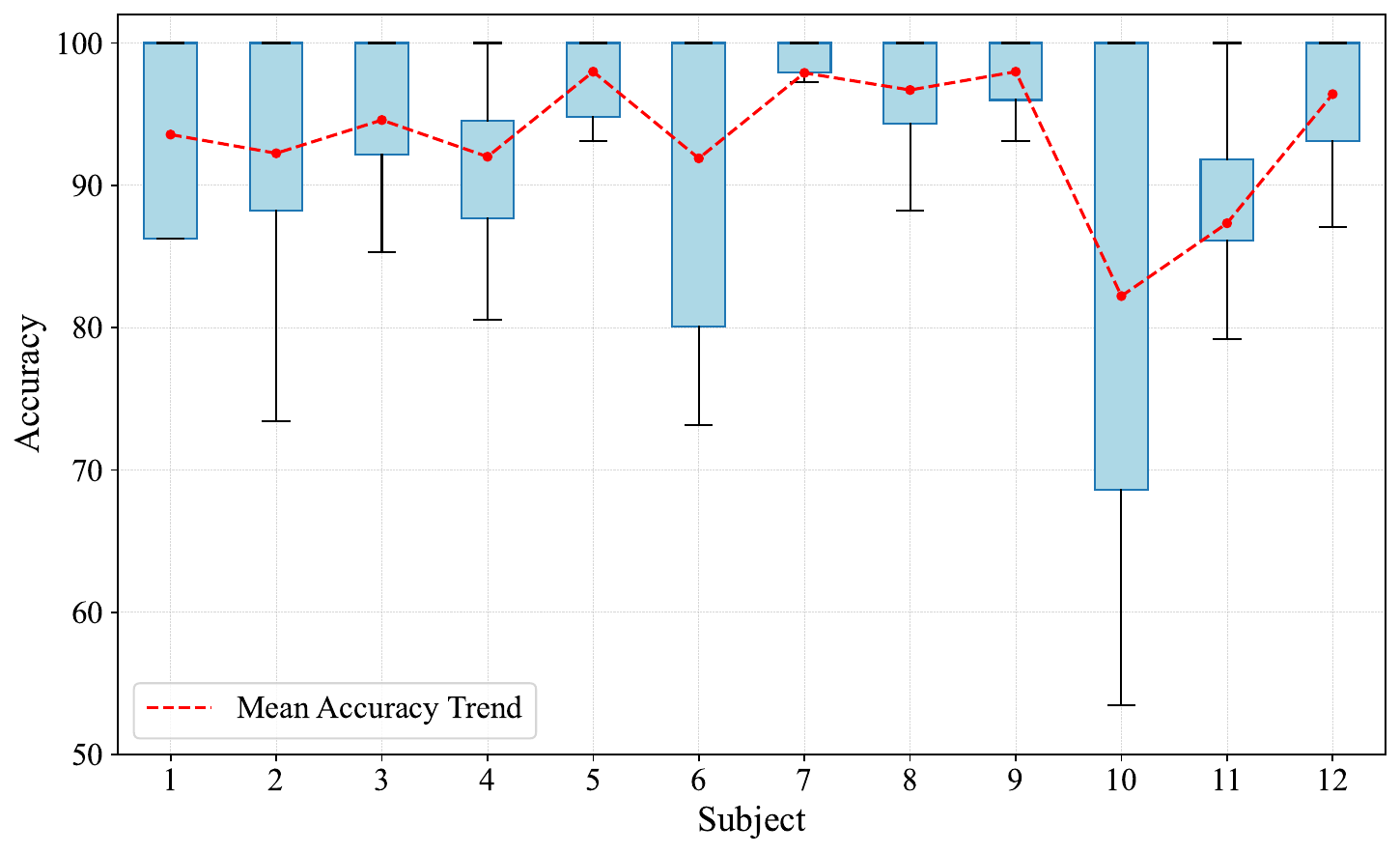}
        \caption{}
    \end{subfigure}
    \hfill 
    \begin{subfigure}[b]{0.35\textwidth}
        \centering
        \includegraphics[height=6cm]{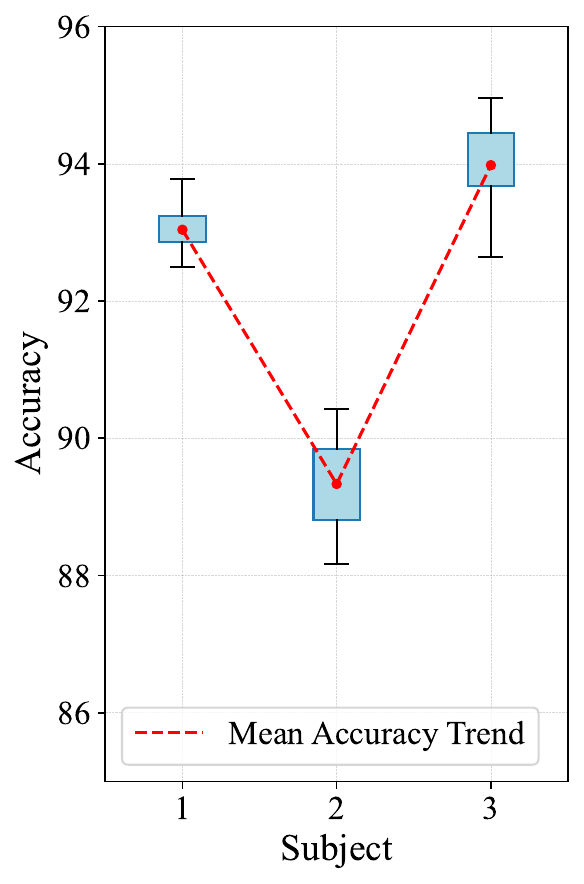}
        \caption{}
    \end{subfigure}
    \caption{Stability analysis of UF-AMA on the SEED dataset under repeated runs with different random seeds: (a) Cross-subject experiments; (b) Cross-session experiments.}
    \label{FIG:4}
\end{figure*}

\begin{figure*}[tbp]
    \centering
    \begin{subfigure}[b]{0.6\textwidth}
        \raggedright
        \includegraphics[height=6cm]{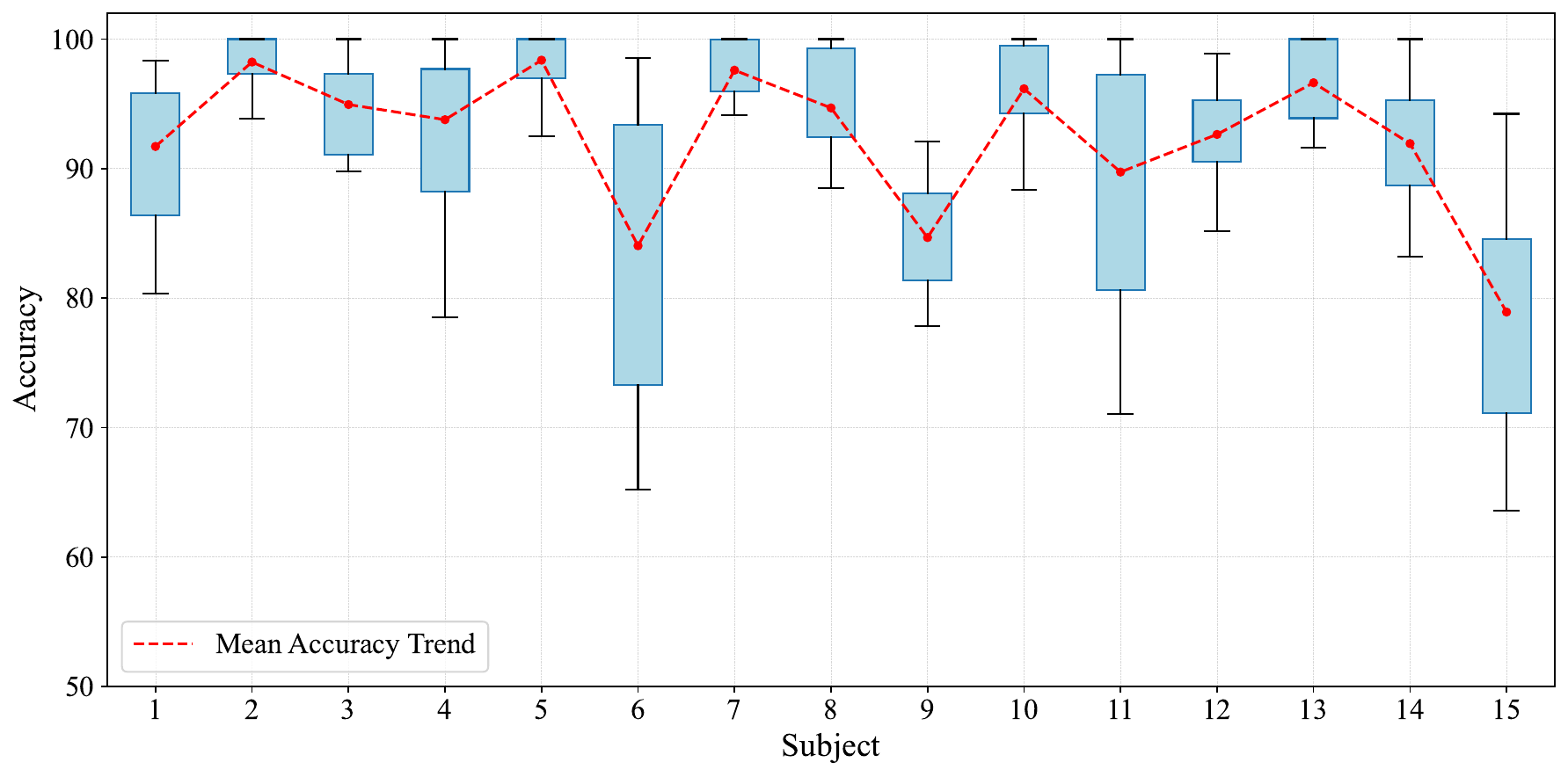}
        \caption{}
    \end{subfigure}
    \hfill
    \begin{subfigure}[b]{0.35\textwidth}
        \centering
        \includegraphics[height=6cm]{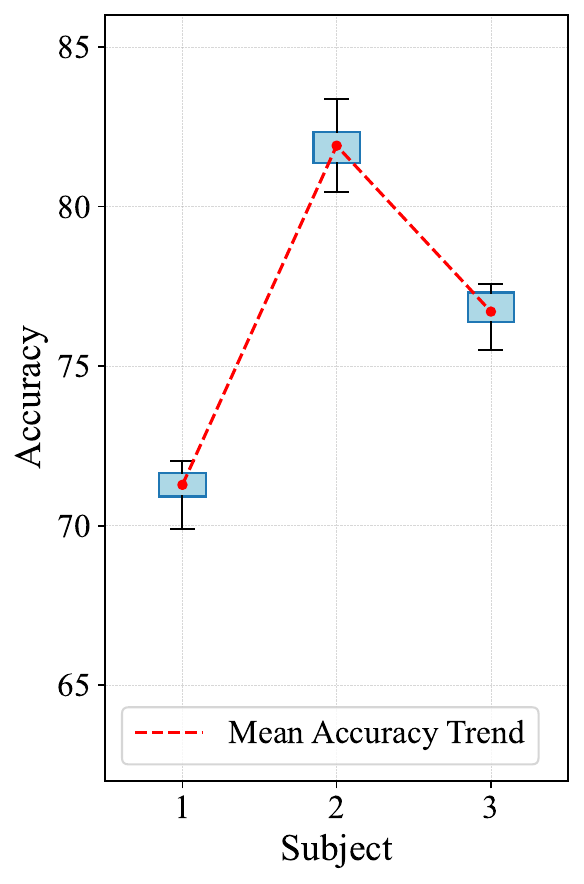}
        \caption{}
    \end{subfigure}
    \caption{Stability analysis of UF-AMA on the SEED-IV dataset under repeated runs with different random seeds: (a) Cross-subject experiments; (b) Cross-session experiments.}
    \label{FIG:5}
\end{figure*}

\begin{figure*}[tbp]
    \centering
    \begin{subfigure}{0.24\textwidth}
        \centering
        \includegraphics[width=\linewidth]{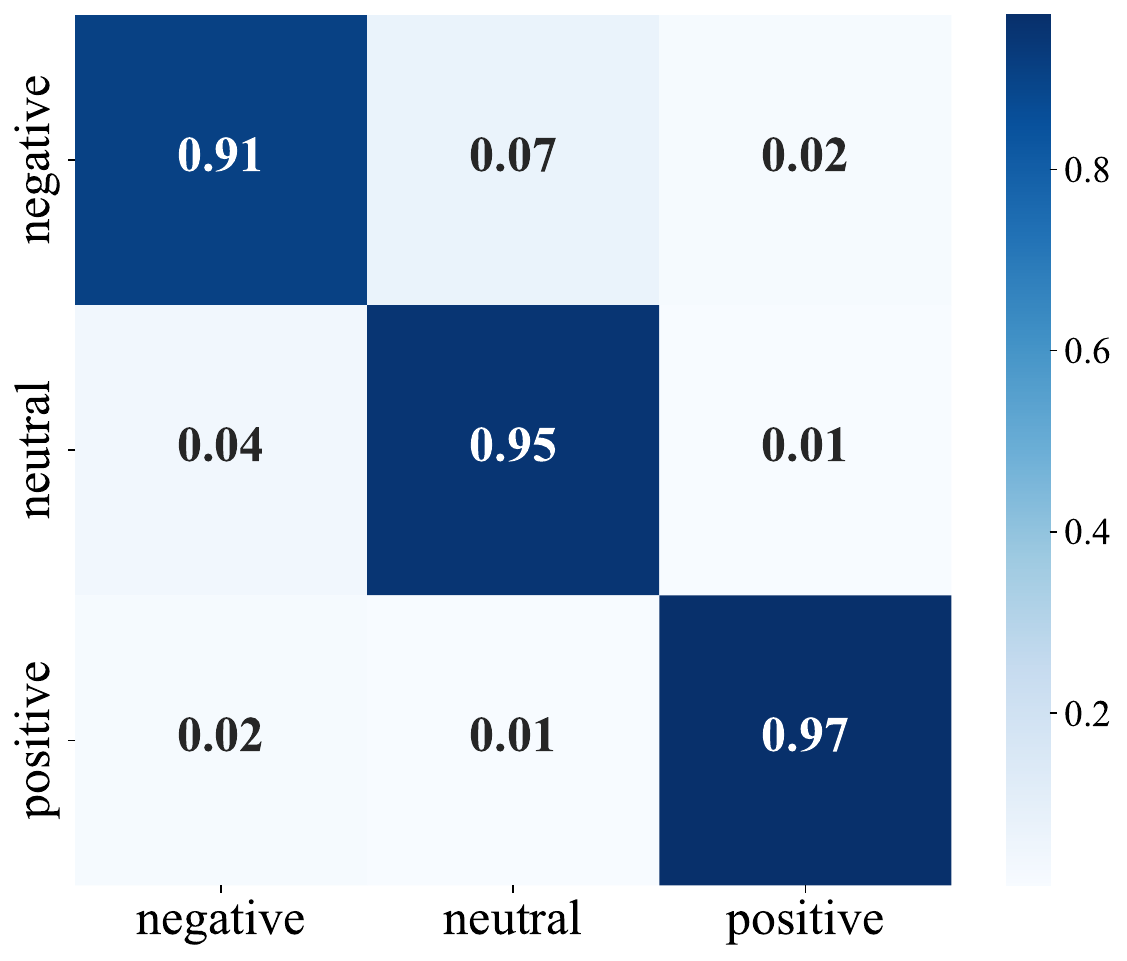}
        \caption{}
        \label{FIG:cm1}
    \end{subfigure}
    \hfill
    \begin{subfigure}{0.24\textwidth}
        \centering
        \includegraphics[width=\linewidth]{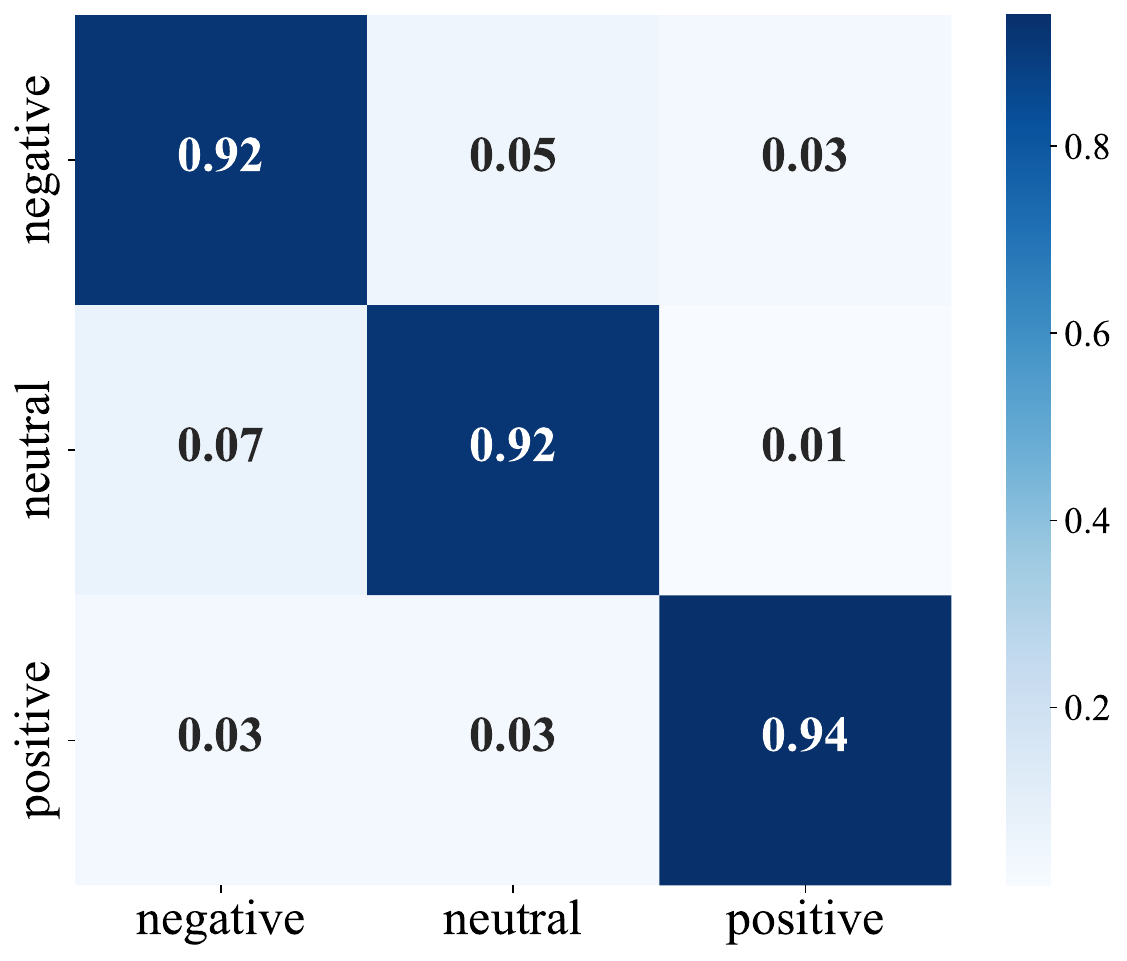}
        \caption{}
        \label{FIG:cm2}
    \end{subfigure}
    \hfill
    \begin{subfigure}{0.24\textwidth}
        \centering
        \includegraphics[width=\linewidth]{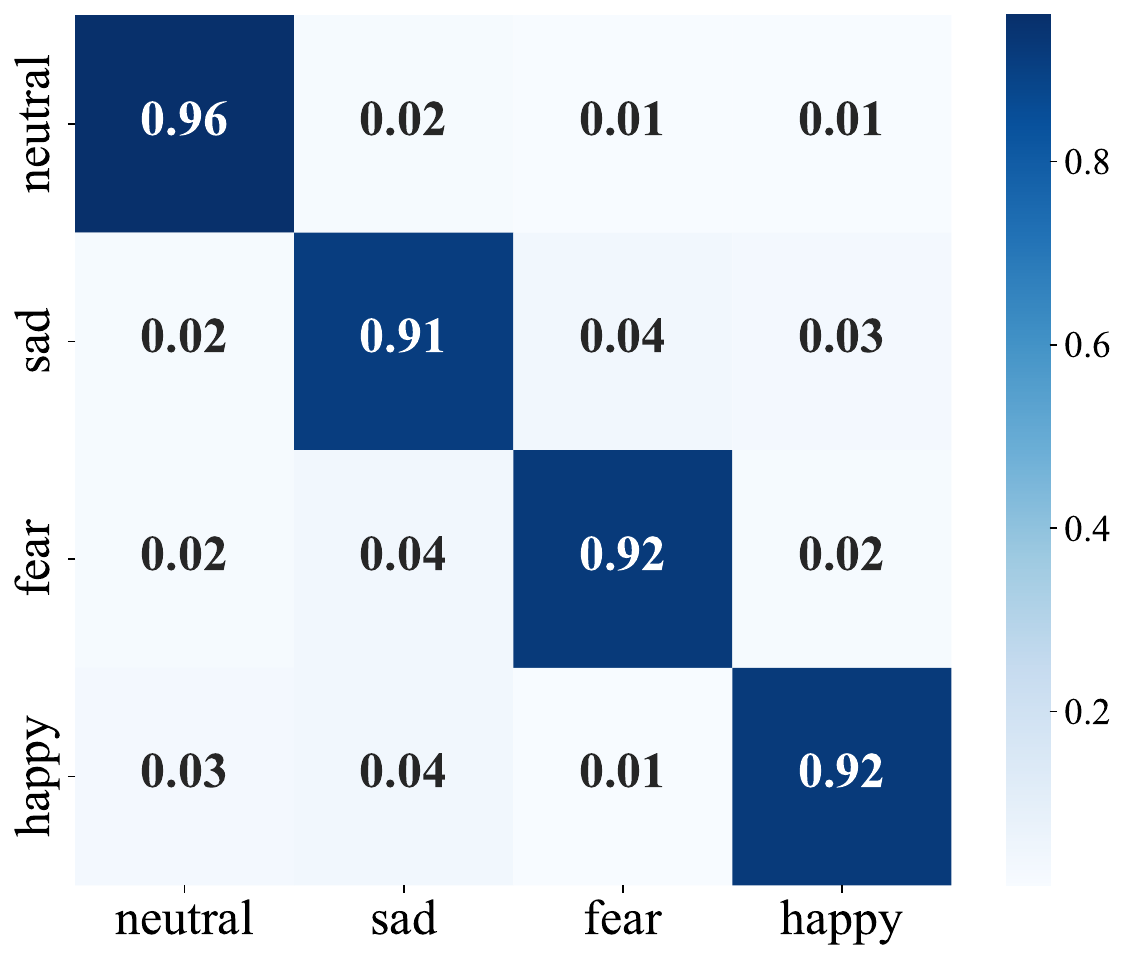}
        \caption{}
        \label{FIG:cm3}
    \end{subfigure}
    \hfill
    \begin{subfigure}{0.24\textwidth}
        \centering
        \includegraphics[width=\linewidth]{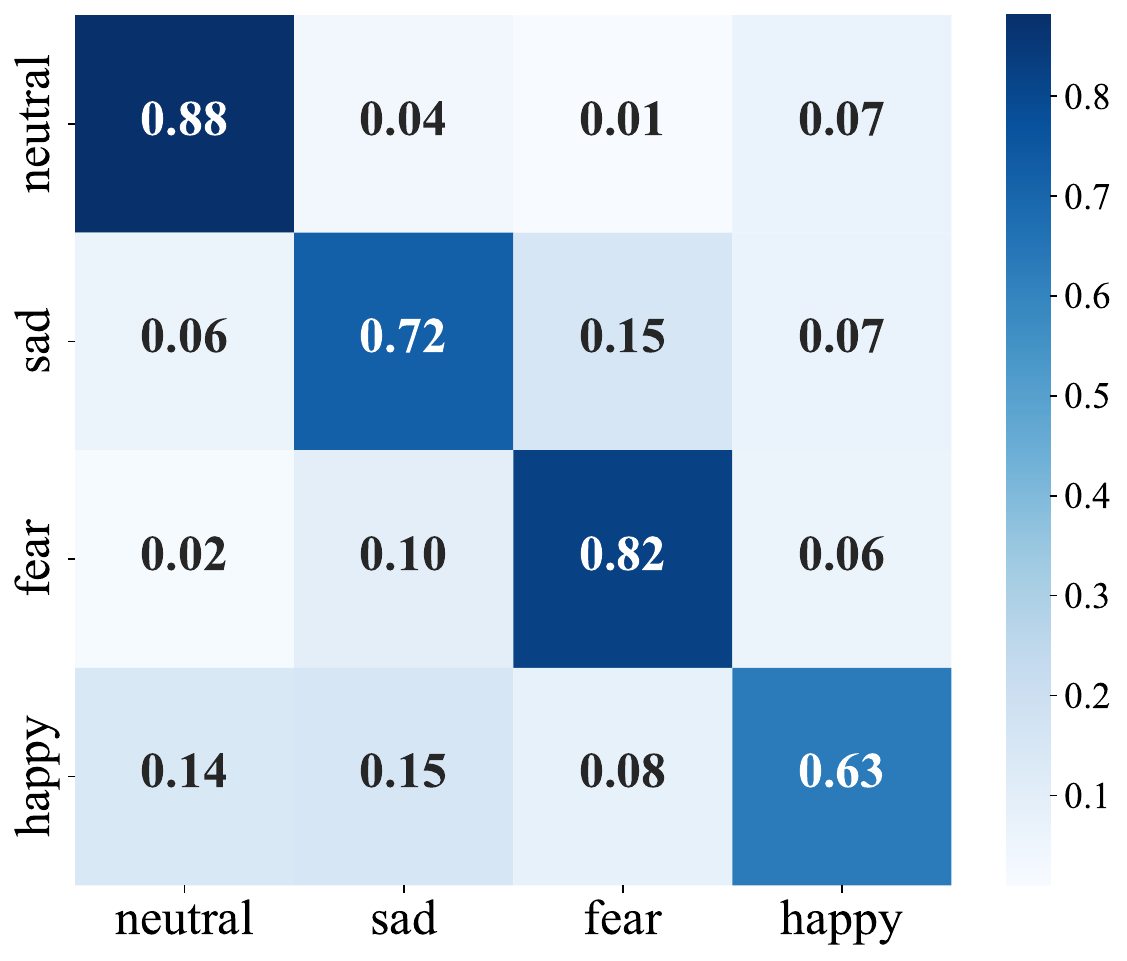}
        \caption{}
        \label{FIG:cm4}
    \end{subfigure}
    \caption{Confusion matrices of UF-AMA on the SEED and SEED-IV datasets under the best-performing seed: (a) Cross-subject experiment on the SEED dataset; (b) Cross-session experiment on the SEED dataset; (c) Cross-subject experiment on the SEED-IV dataset; (d) Cross-session experiment on the SEED-IV dataset. The diagonal values represent the classification accuracy for each emotion category.}
    \label{FIG:6}
\end{figure*}

To evaluate the stability of UF-AMA, we ran multiple repeated experiments with different random seeds on the SEED dataset. As shown in \cref{FIG:4}, the average accuracy for the majority of subjects in the cross-subject experiment remained at a high level, with the interquartile range (IQR) of the box plots around 10\%. Among them, Subject 6 and Subject 10 exhibited greater variability. The IQR of their box plots was larger, indicating that the model was more sensitive to the choice of random seeds for these individuals. In contrast, the cross-session experiment demonstrated substantially higher stability. The average accuracy for all three sessions remained above 90\%, and the IQR of the box plots was within 5\%, indicating high reproducibility of the model's performance across sessions. The performance fluctuation under different seeds was greater in the cross-subject experiment, which may be attributed to the greater variance in data distribution across subjects.

\cref{FIG:6} presents the confusion matrices obtained from the cross-subject and cross-session experiments on the SEED dataset, both plotted based on the results under the best-performing seed for each experiment. The diagonals of the confusion matrices for both types of experiments exhibit darker colors, indicating that UF-AMA achieved good emotion classification performance in both cross-subject and cross-session settings. As shown in \cref{FIG:cm1}, the model in the cross-subject experiment achieved over 90\% accuracy for all three emotion recognition tasks, with accuracies of 0.91, 0.95, and 0.97 for the negative, neutral, and positive labels, respectively. As shown in \cref{FIG:cm2}, the model in the cross-session experiment also achieved over 90\% accuracy for all three emotion recognition tasks, with accuracies of 0.92, 0.92, and 0.94 for the negative, neutral, and positive labels, respectively.

\subsection{Experiments on the SEED-IV dataset}

We evaluated UF-AMA on the SEED-IV dataset under the same protocols, and systematically compared the proposed method against a series of state-of-the-art baseline models. In addition to the previously mentioned DA-CapsNet \cite{Liu2024}, PR-PL \cite{Zhou2023}, TSSSA-Net \cite{Guo2025}, LGDAAN-Nets \cite{An2025a}, SDC-Net \cite{Tang2025}, FMLAN \cite{Yu2025}, CFDA-CSF \cite{Jimenez-Guarneros2024a}, and LAFDA-Net \cite{Xu2024}, the compared baseline models also included: RHPR-Net \cite{Tang2024}, which employs spatial-frequency pattern extractors to recognize EEG dual-domain representations and utilizes a hierarchical module to integrate heterogeneous central and peripheral features; MMDA \cite{Jimenez-Guarneros2024b}, which performs dual alignment of marginal and conditional distributions and introduces a prioritization strategy to suppress negative transfer in cross-subject transfer; CSMM \cite{Zhu2025b}, which utilizes dynamic adversarial domain adaptation for cross-subject alignment and combines attention mechanisms with contrastive loss; and C2PCI-Net \cite{Ma2025}, which employs hybrid encoders to unify EEG channel mapping and embeds adjacent cross-modal information from peripheral signals to achieve central-peripheral complementary integration.

\begin{table}[htbp]
    \centering
    \captionsetup{singlelinecheck=off, font={bf, small}, labelsep=period}
    \caption{Performance comparison of various recognition methods in terms of accuracy and standard deviation on the SEED-IV dataset.}
    \label{TAB:6}
    \setlength{\tabcolsep}{6pt}
    \footnotesize
    \renewcommand{\arraystretch}{1.3} 
    \begin{tabular*}{\columnwidth}{@{\hspace{6pt}} l l c c @{\hspace{6pt}}}
        \toprule
        Years & Method & \multicolumn{2}{c}{Accuracy/Std (\%)} \\ 
        \cmidrule(lr){3-4} & & Cross-subject & Cross-session \\
        \midrule
        2024 & DA-CapsNet \cite{Liu2024}   & 70.84 $\pm$ 09.35 & 75.38 $\pm$ 06.19 \\
        2024 & PR-PL \cite{Zhou2023}       & 74.92 $\pm$ 08.53 & 74.62 $\pm$ 14.15 \\
        2025 & TSSSA-Net \cite{Guo2025}    & 74.02 $\pm$ 10.25 & -- \\
        2025 & LGDAAN-Nets \cite{An2025a}  & 74.12 $\pm$ 08.45 & 76.70 $\pm$ 08.16 \\
        2025 & SDC-Net \cite{Tang2025}     & 74.88 $\pm$ 10.47 & -- \\
        2025 & FMLAN \cite{Yu2025}         & 76.01 $\pm$ 08.50 & \textbf{78.78 $\pm$ 10.78} \\
        2024 & RHPRNet \cite{Tang2024}     & 67.62 $\pm$ 07.26 & -- \\
        2024 & C2PCI-Net \cite{Ma2025}     & 71.94 $\pm$ 08.80 & -- \\
        2024 & CFDA-CSF \cite{Jimenez-Guarneros2024a}   & 87.40 $\pm$ 09.90 & -- \\
        2024 & LAFDA-Net \cite{Xu2024}  & \underline{89.85 $\pm$ 01.04} & -- \\
        2025 & MMDA \cite{Jimenez-Guarneros2024b}       & 85.54 $\pm$ 08.11 & -- \\
        2025 & CSMM \cite{Zhu2025b}        & 89.82 $\pm$ 06.22 & -- \\
        \midrule
        2026 & UF-AMA & \textbf{92.73 $\pm$ 07.27} & \underline{77.24 $\pm$ 04.81} \\
        \bottomrule
    \end{tabular*}
    \flushleft{\scriptsize The best experimental results are bolded, and the second-best experimental results are underlined.}
\end{table}

\begin{table}[htbp]
    \centering
    \captionsetup{singlelinecheck=off, font={bf, small}, labelsep=period}
    \caption{Ablation study of different components in the cross-subject experiment in terms of accuracy and standard deviation on the SEED and SEED-IV datasets.}
    \label{TAB:7}
    \setlength{\tabcolsep}{0pt} 
    \footnotesize
    \renewcommand{\arraystretch}{1.3} 
    \begin{tabular*}{\columnwidth}{@{\hspace{8pt}} l @{\extracolsep{\fill}} c c @{\hspace{8pt}}} 
        \toprule
        Method & \multicolumn{2}{c}{Accuracy/Std (\%)} \\ 
        \cmidrule{2-3} & SEED & SEED-IV \\ 
        \midrule
        Model 1 & 91.85 $\pm$ 08.59 & 91.29 $\pm$ 07.41 \\
        Model 2 & 92.04 $\pm$ 08.21 & 90.41 $\pm$ 07.41 \\
        Model 3 & 93.79 $\pm$ 06.70 & 92.40 $\pm$ 06.68 \\
        Model 4 & 93.41 $\pm$ 07.27 & 92.41 $\pm$ 07.90 \\
        Model 5 & 94.48 $\pm$ 06.06 & 91.76 $\pm$ 07.89 \\
        Model 6 & \textbf{94.53 $\pm$ 07.11} & \textbf{92.73 $\pm$ 07.27} \\
        \bottomrule
    \end{tabular*}
\end{table}

\cref{TAB:6} presents the performance comparison of different methods on the SEED-IV dataset. The experimental results show that the proposed UF-AMA achieved an accuracy of 92.73\% in the cross-subject experiment, ranking first among all compared algorithms. Meanwhile, it achieved an accuracy of 77.24\% in the cross-session experiment, ranking second among all algorithms, surpassed only by FMLAN (78.78\%). Compared with the baseline models, UF-AMA demonstrated clear performance advantages in both cross-subject and cross-session experiments. In the cross-subject experiment, its accuracy was 2.88\% higher than that of LAFDA-Net (89.85\%), a model with prominent performance in multimodal tasks, and 2.91\% higher than that of the equally competitive CSMM (89.82\%). In the cross-session experiment, its accuracy was 0.54\% higher than that of LGDAAN-Nets (76.70\%). In the cross-subject experiment, while achieving the highest accuracy, UF-AMA also exhibited a standard deviation superior to most high-performance baseline models. In the cross-session experiment, although the average accuracy of UF-AMA was 1.54\% lower than that of FMLAN, its standard deviation was merely 4.81, significantly lower than the latter's 10.78.

As shown in \cref{TAB:5}, the proposed UF-AMA also demonstrated strong performance across multiple sessions of the SEED-IV dataset. UF-AMA achieved the highest accuracy among all compared methods on Subject 1, Subject 2, Subjects 5–7, Subject 10, Subject 12, and Subject 14 in Session 1. In both Session 2 and Session 3, UF-AMA achieved the best performance on 13 subjects. Across all subjects, UF-AMA achieved an average accuracy of 90.66\% in Session 1, representing improvements of 4.94\% and 1.67\% over CFDA-CSF and LAFDA-Net, respectively. In Session 2, UF-AMA achieved an average accuracy of 94.66\%, representing improvements of 5.06\% and 3.77\% over CFDA-CSF and LAFDA-Net, respectively. In Session 3, UF-AMA achieved an average accuracy of 92.86\%, representing improvements of 5.98\% and 4.39\% over CFDA-CSF and LAFDA-Net, respectively. Furthermore, the standard deviations of UF-AMA in Session 1, Session 2, and Session 3 were only 7.32, 4.13, and 8.93, respectively, which were significantly lower than the corresponding 11.02, 6.65, and 10.79 for CFDA-CSF and 10.40, 6.36, and 9.47 for LAFDA-Net. The experimental results demonstrate that even in a four-class emotion recognition task with more ambiguous category boundaries and more complex feature distributions, UF-AMA can still consistently and effectively suppress interference caused by signal fluctuations, exhibiting strong generalization ability and reliability.

We ran multiple repeated experiments with different random seeds on the SEED-IV dataset. As shown in \cref{FIG:5}, the cross-subject experiment exhibited substantial individual variability. Although the average accuracy for the majority of subjects remained at a high level, the IQR of the box plots varied considerably across subjects. In contrast, the cross-session experiment demonstrated higher stability. Although the average accuracy for the three sessions ranged between 70\% and 85\% due to task difficulty, the IQR for each session was less than 5\%, indicating high reproducibility of the model's performance across sessions.

\cref{FIG:cm3} and \cref{FIG:cm4} present the confusion matrices for the cross-subject and cross-session experiments on the SEED-IV dataset. In the cross-subject experiment, UF-AMA demonstrated strong discriminative ability, achieving accuracies of 0.96, 0.91, 0.92, and 0.92 for the neutral, sad, fear, and happy labels, respectively, with a deeply colored and uniformly distributed diagonal. However, in the cross-session experiment, while the model maintained a high recognition rate of 0.88 for the neutral label, the accuracy for the happy label decreased, with nearly 15\% of the samples misclassified as neutral and a similar proportion as sad. Additionally, some samples under the sad label were misclassified as fear. A plausible explanation is that, affected by the time interval between sessions, the subjects' physiological arousal levels may have weakened due to state fluctuations, causing a shift in the arousal level associated with happiness. Moreover, sadness and fear are both negative emotions and are relatively close in arousal level. This phenomenon can be further amplified under cross-session fluctuations, making it difficult for the model to accurately distinguish between these negative emotions.

\subsection{Ablation study}
\label{SUBSEC:3.5}

\begin{figure*}[tbp]
    \centering
    \begin{subfigure}{0.49\textwidth}
        \centering
        \includegraphics[width=1.0\linewidth]{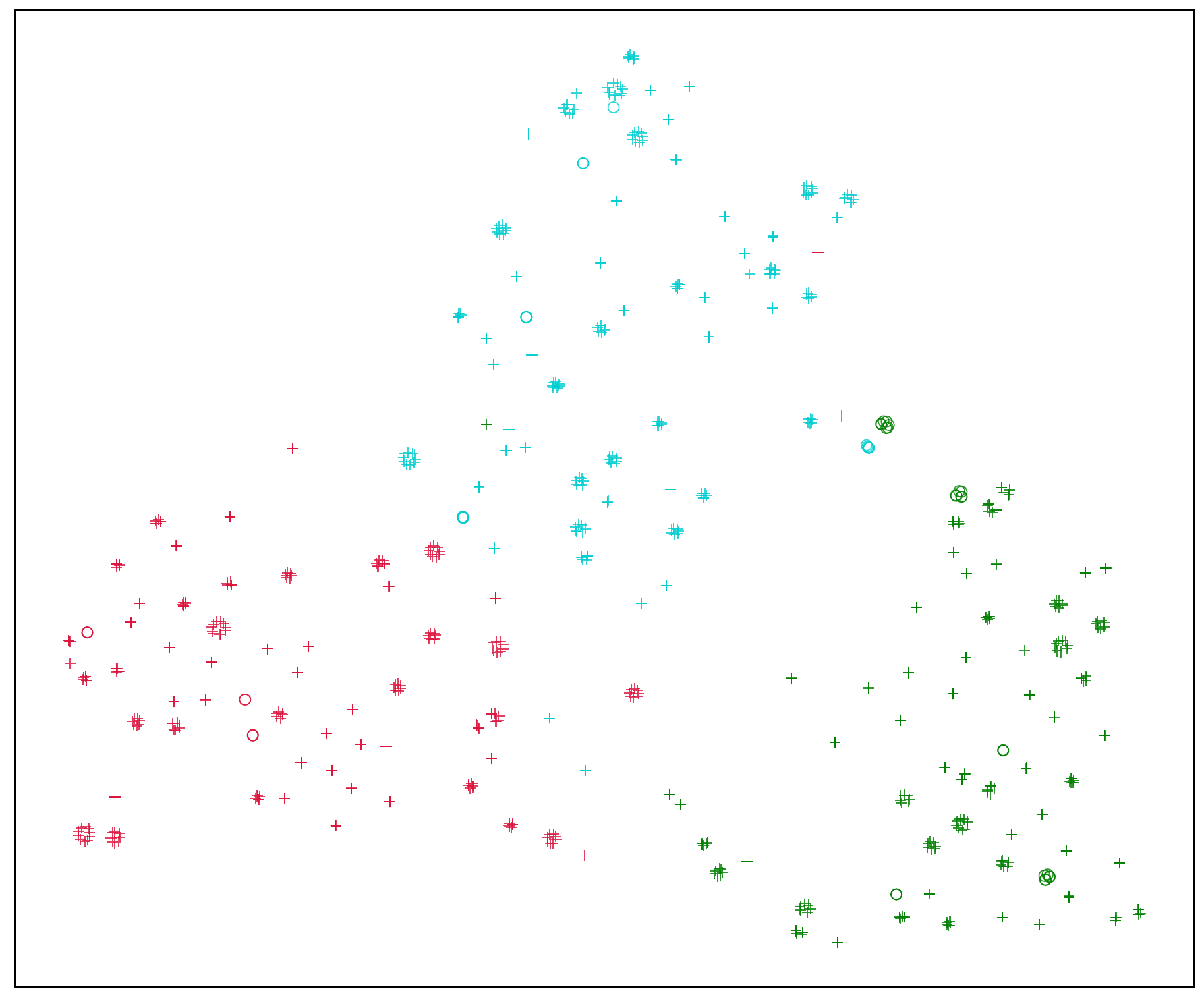}
        \caption{}
        \label{FIG:tsne1}
    \end{subfigure}
    \hfill
    \begin{subfigure}{0.49\textwidth}
        \centering
        \includegraphics[width=1.0\linewidth]{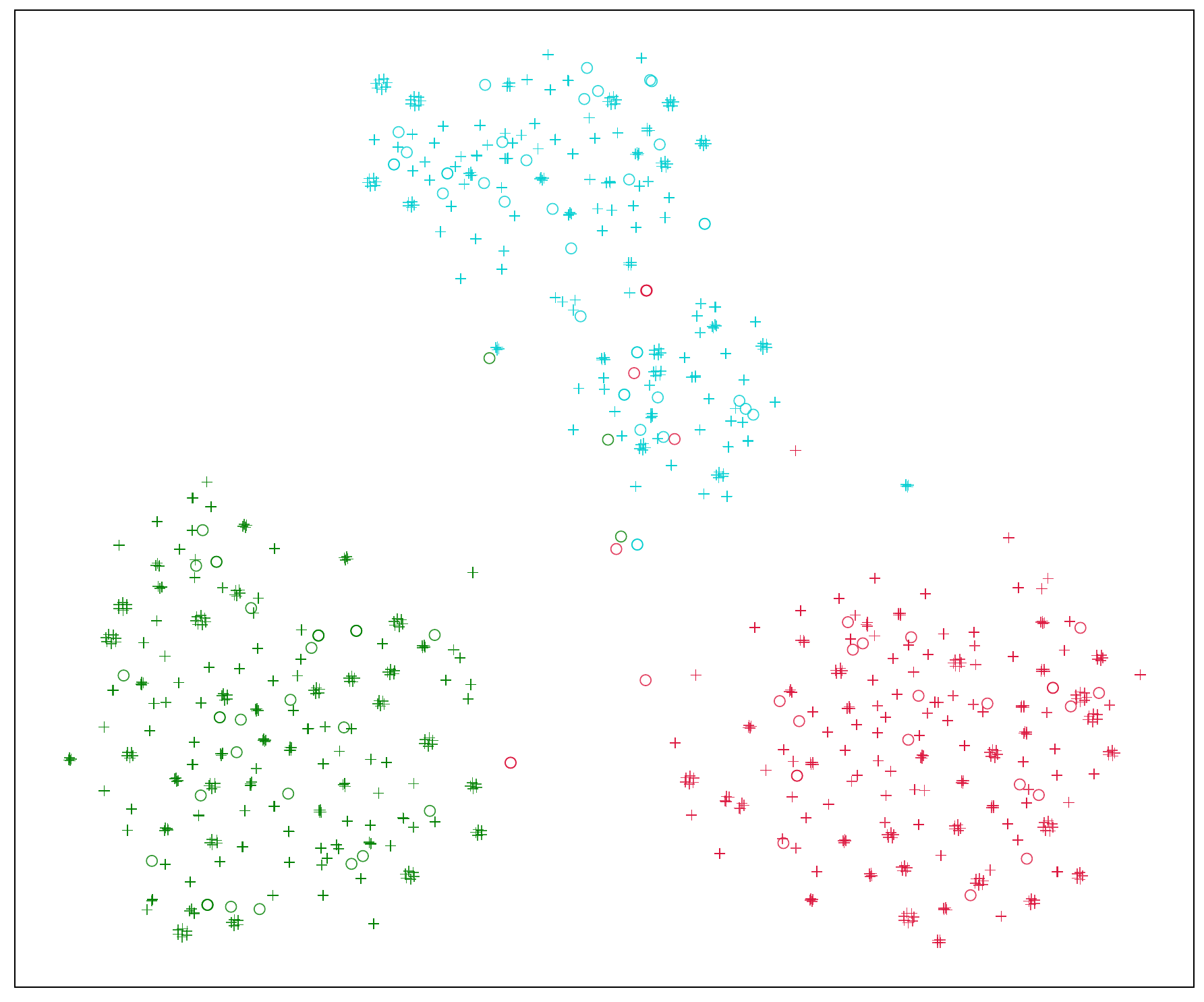}
        \caption{}
        \label{FIG:tsne2}
    \end{subfigure}
    \vspace{10pt}
    \begin{subfigure}{0.49\textwidth}
        \centering
        \includegraphics[width=1.0\linewidth]{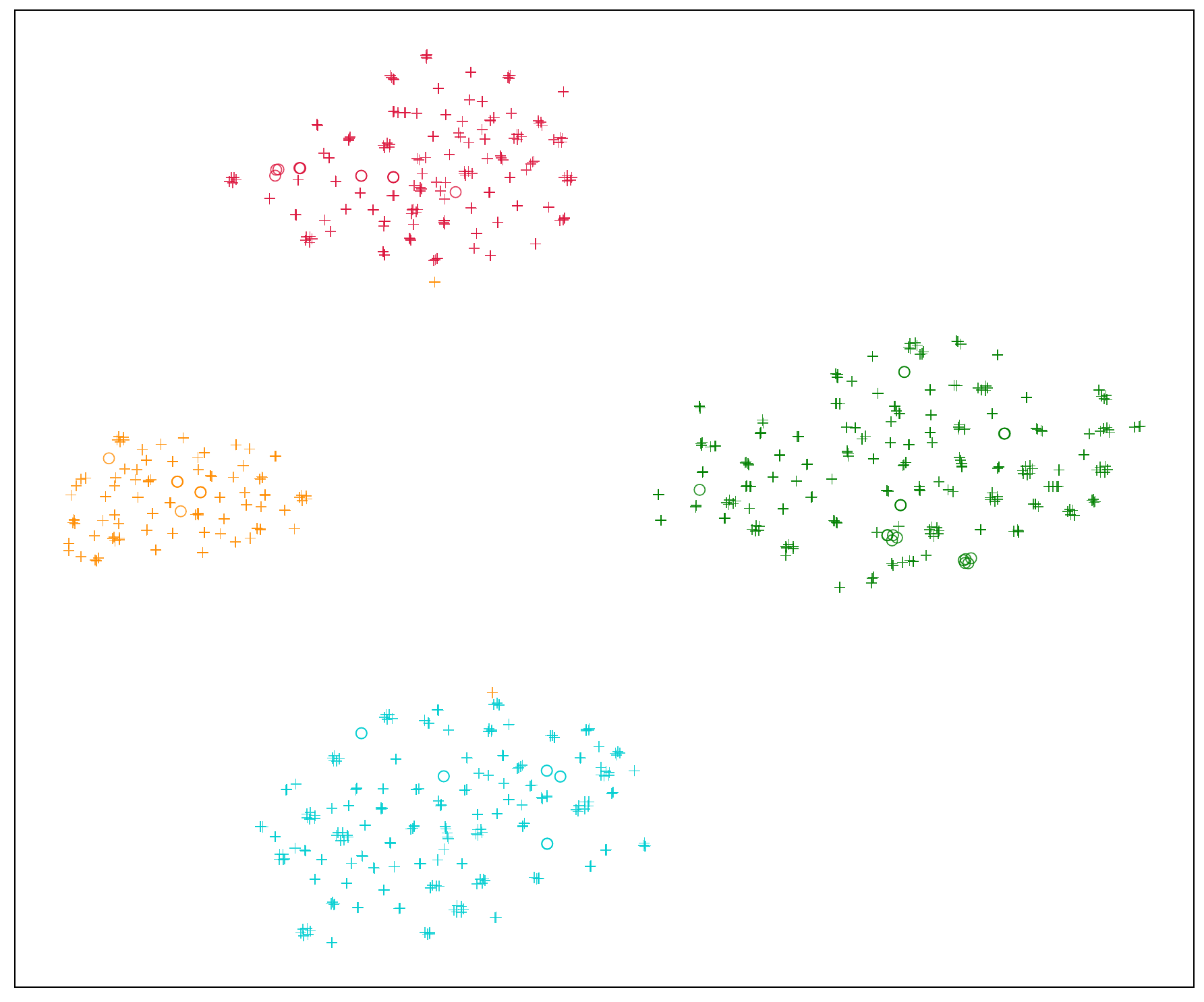}
        \caption{}
        \label{FIG:tsne3}
    \end{subfigure}
    \hfill
    \begin{subfigure}{0.49\textwidth}
        \centering
        \includegraphics[width=1.0\linewidth]{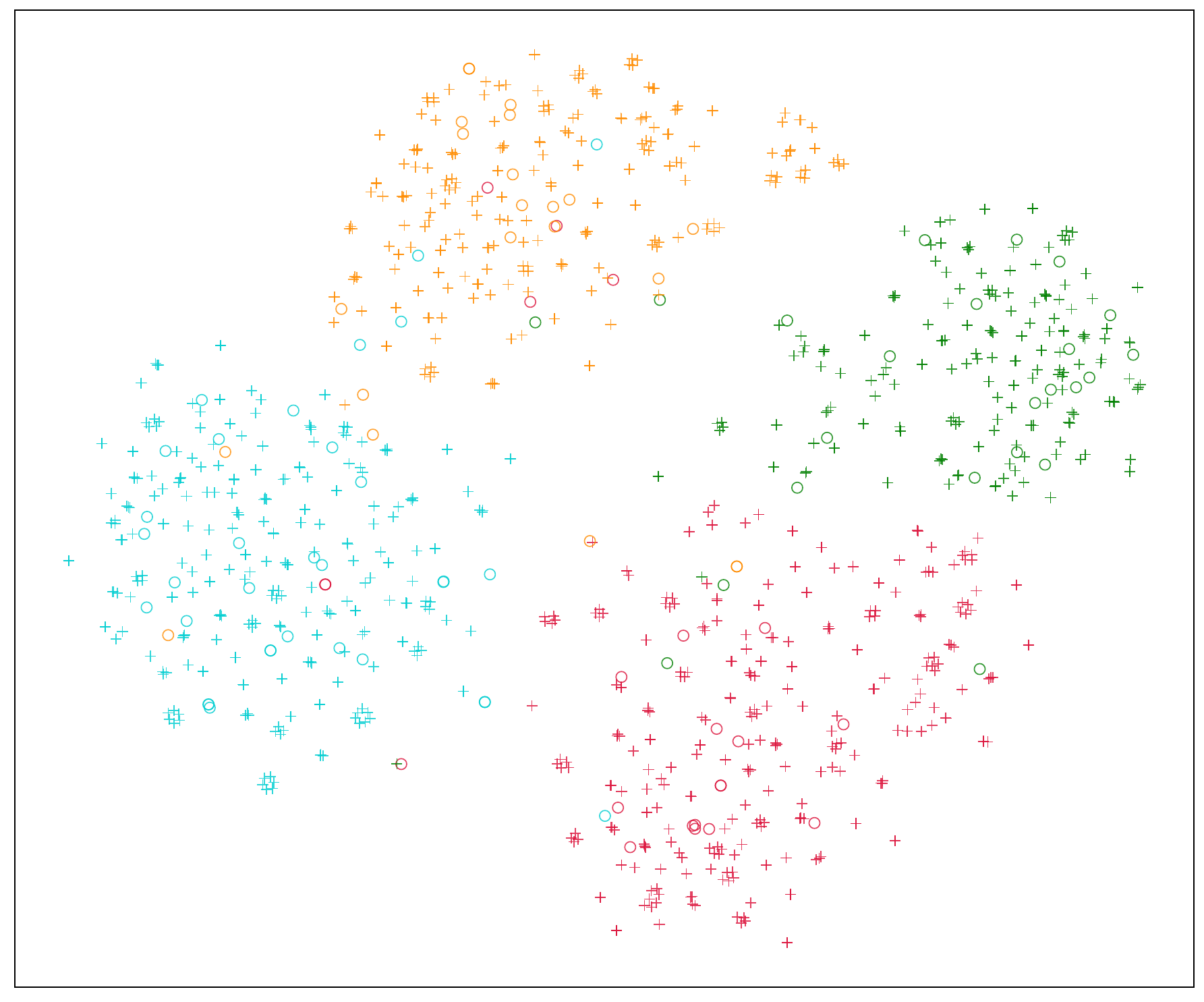}
        \caption{}
        \label{FIG:tsne4}
    \end{subfigure}
    \caption{t-SNE feature visualization of UF-AMA on the SEED and SEED-IV datasets: (a) Cross-subject experiment on the SEED dataset (Subject 5, Session 3); (b) Cross-session experiment on the SEED dataset (Session 3); (c) Cross-subject experiment on the SEED-IV dataset (Subject 8, Session 2); (d) Cross-session experiment on the SEED-IV dataset (Session 2).}
    \label{FIG:7}
\end{figure*}

To systematically evaluate the contribution of each functional module within UF-AMA, we constructed six variant models using an incremental strategy for the ablation study: (1) Model 1: the cross-modal feature fusion network trained solely on the source domain dataset, without any additional loss or alignment modules; (2) Model 2: adds intra-class compactness loss based on binary cross-entropy to Model 1; (3) Model 3: adds local-global coarse-grained marginal alignment via MMD to Model 2; (4) Model 4: introduces global consistency alignment applied to the target domain data on top of Model 3, without employing any sample selection mechanism; (5) Model 5: introduces the confidence-based high-quality target data selection mechanism on top of Model 4; (6) Model 6: integrates local cross-modal distillation on top of Model 5, forming the complete UF-AMA. We conducted cross-subject experiments for these six models on the SEED and SEED-IV datasets using the experimental strategy described in \cref{SUBSEC:3.2}. The detailed experimental results are presented in \cref{TAB:7}.

As shown in \cref{TAB:7}, the incremental introduction of each module brought cumulative performance improvements. The result of Model 2 slightly outperforming Model 1 indicates that enforcing intra-class clustering based on semantic consistency in the source domain helps refine the feature extraction process. The substantial accuracy improvement of Model 3 upon the introduction of local-global coarse-grained alignment validates the effectiveness of structured marginal alignment in mitigating cross-domain distribution shifts. Notably, although Model 4 introduced global consistency alignment, its gains were relatively limited and, in some tasks, even suffered from negative interference due to the lack of constraints on pseudo-label noise. For Model 5, which introduced the confidence-based high-quality target domain sample selection mechanism on top of Model 4, its accuracy on the SEED dataset improved substantially. This strongly demonstrates the effectiveness of the selection mechanism in filtering noisy data and providing high-quality alignment guidance. However, on the SEED-IV dataset, the model performance experienced a slight drop, which may be attributed to the selection mechanism mistakenly discarding some more representative target domain samples, thereby leading to reduced generalization ability. The final Model 6, as the complete version of UF-AMA, achieved the highest accuracy on both the SEED and SEED-IV datasets, thereby validating that local cross-modal distillation allows the model to extract informative cross-modal cues from EEG signals and eye movement data even in the presence of noise, thereby enhancing its performance in complex emotion recognition tasks.

Furthermore, the incremental introduction of each module also enhanced the stability of the model output. As the alignment mechanisms and selection strategy were refined, the model's performance fluctuation gradually converged, with the standard deviation showing an overall decreasing trend from Model 1 to UF-AMA. This indicates that each module contributes to the model's robustness against interference such as individual differences among subjects, endowing the model with greater reliability in complex cross-domain emotion recognition tasks.

\subsection{Visualization}

To qualitatively evaluate the effectiveness of UF-AMA, we selected specific subjects and sessions as the target domain for the cross-subject and cross-session experiments on both the SEED and SEED-IV datasets. After model training, we performed t-SNE feature visualization analysis \cite{VanDerMaaten2008} for the four models. As shown in \cref{FIG:7}, different colors correspond to different emotion categories, and the symbols $+$ and $\bigcirc$ represent the source domain data and the target domain data, respectively. \cref{FIG:tsne1} corresponds to the cross-subject experiment with Subject 5 as the target domain in Session 3 of the SEED dataset; \cref{FIG:tsne2} corresponds to the cross-session experiment with Session 3 as the target domain in the SEED dataset; \cref{FIG:tsne3} corresponds to the cross-subject experiment with Subject 8 as the target domain in Session 2 of the SEED-IV dataset; and \cref{FIG:tsne4} corresponds to the cross-session experiment with Session 2 as the target domain in the SEED-IV dataset.

As observed from the visualization results, UF-AMA demonstrates excellent deep feature representation capability and domain alignment robustness across different task scenarios. In both the cross-subject and cross-session experiments on the SEED dataset, the three emotion features, after being extracted by the model, exhibited clear inter-class separability, and the source and target domain distributions achieved a high degree of alignment, indicating that the model can effectively capture domain-invariant emotional representations. In the more complex emotion categories of the cross-subject and cross-session experiments on the SEED-IV dataset, the four emotion features still exhibited clear inter-class separability and intra-class compactness. Moreover, the majority of target domain samples were mapped to the central distribution area of the source domain features of the same class, without noticeable category overlap or boundary ambiguity. This further indicates that UF-AMA possesses strong classification discriminability and generalization performance in complex tasks.

\section{Discussion}

In this study, we proposed UF-AMA, an emotion recognition framework based on multimodal physiological signals. Within this framework, we constructed a local-global collaborative feature extraction and alignment architecture, and introduced a confidence-guided mechanism designed to dynamically perceive and handle modality quality discrepancies in cross-domain scenarios, thus offering a novel perspective for multimodal emotion recognition research. Building upon this, we designed a multi-granularity domain adaptation structure that integrates marginal and conditional distributions to achieve distributional consistency and semantic alignment at different granularities. We extensively evaluated the proposed method on the SEED and SEED-IV datasets. As shown in \cref{TAB:3} and \cref{TAB:6}, unimodal emotion recognition methods like ACRNN are limited by the singularity of physiological representations. This makes it difficult for them to cope with the non-stationarity of physiological signals and individual differences, resulting in relatively poor overall performance. Although some existing multimodal methods enhance performance by integrating heterogeneous complementary information, during the cross-domain transfer process, these methods often rely excessively on the global feature space alignment of the fusion branch, thereby masking the unique distributional biases of individual modality branches. Furthermore, lacking semantically guided conditional distribution alignment, they struggle to fully exploit the fine-grained category boundary information embedded in the target domain pseudo-labels. In addition, even though some studies like SDC-Net attempt to introduce conditional distribution alignment, they can easily overlook the quality differences among different modalities of data in the target domain, making the model susceptible to interference from low-quality modality noise during the alignment process.

Therefore, we proposed UF-AMA to effectively address the aforementioned challenges. The comparative results in \cref{TAB:3} and \cref{TAB:6} demonstrate that UF-AMA achieves superior recognition performance on both the SEED and SEED-IV datasets compared to current state-of-the-art methods. \cref{TAB:4} and \cref{TAB:5} specifically detail the model performance for each subject across different sessions in the cross-subject experiments. The results show that despite significant variability among individuals, UF-AMA attained the optimal recognition accuracy for the vast majority of subject-session combinations. To further investigate the contributions of individual modules, we conducted an incremental ablation study. The experimental results in \cref{TAB:7} confirm that the local-global collaborative alignment mechanism plays a crucial role in mitigating domain shift. Meanwhile, the confidence-based sample selection module optimizes the process of deep mining cross-modal fine-grained semantic information by filtering pseudo-label noise, enabling a substantial performance gain. Furthermore, the visualization results in \cref{FIG:7} clearly demonstrate the superior distribution alignment capability of UF-AMA.

Despite the strong performance of UF-AMA in cross-domain emotion recognition, this study still has certain limitations. During training, the current framework employs a relatively fixed allocation of loss weights among the modality-specific branches and the fusion branch, which limits the model's flexibility in automatically adjusting modality importance. Additionally, the confidence ranking threshold $\alpha$ currently relies on a preset static parameter, which similarly fails to achieve dynamic adaptive regulation. This limitation was reflected in the SEED-IV ablation study in \cref{SUBSEC:3.5}, where the static selection mechanism may have caused some potentially representative target domain samples to be mistakenly identified as noise and discarded, leading to performance fluctuation. Therefore, in future work, we plan to introduce an adaptive weight allocation strategy and a dynamic threshold adjustment mechanism to optimize the existing framework, aiming to enhance emotion recognition performance and further facilitate its deployment and application in practical tasks.

\section{Conclusion}
In this study, we proposed UF-AMA, an emotion recognition framework based on multimodal physiological signals. Compared with existing state-of-the-art models, our proposed method is not only capable of adaptively addressing modality quality discrepancies through a confidence-aware mechanism, but also integrates multi-granularity information through a local-global collaborative learning strategy, thereby achieving more robust domain alignment. We evaluated UF-AMA on the SEED and SEED-IV datasets, and the results demonstrate that this framework exhibits strong performance in terms of recognition accuracy, generalization ability, and output stability in both cross-subject and cross-session scenarios. UF-AMA not only provides an effective technical pathway for noise suppression and sample quality assessment in multimodal feature fusion but also offers an important reference for constructing highly reliable and robust affective computing systems in complex real-world environments.

\printcredits

\section*{Acknowledgments}
This research was supported by the Yangtze River Delta Science and Technology Innovation Community Joint Research Project under Grant 2025CSJZN01600.

\bibliographystyle{new-style}

\bibliography{new-refs}

\end{document}